\documentclass[preprint, 5p, times, sort&compress]{elsarticle}

\usepackage{amsfonts, amsmath, amssymb}
\numberwithin{equation}{section}

\usepackage[hidelinks, pdfauthor=author]{hyperref}

\usepackage{verbatim, xcolor}
\usepackage{tabularx, enumitem, epsfig, float, subcaption, cleveref}
\usepackage[normalem]{ulem}

\usepackage{pgfplots, tikz, tikzscale}
\pgfplotsset{compat=1.14}

\newcommand{\subf}[2]{{\small \begin{tabular}[t]{@{}c@{}} #1\\#2 \end{tabular}}}

\let\vec\boldsymbol 

\begin{document}
\begin{frontmatter}
\journal{}

\title{Systematic two-scale image analysis of extreme deformations in soft architectured sheets}

\author[address1]{Filippo Agnelli}
\author[address1]{Pierre Margerit}
\author[address2]{Paolo Celli}
\author[address3]{Chiara Daraio}
\author[address1]{Andrei Constantinescu\corref{mycorrespondingauthor}}
\cortext[mycorrespondingauthor]{Corresponding author}
\ead{andrei.constantinescu@polytechnique.edu}

\address[address1]{Laboratoire de Mécanique des Solides, CNRS, École polytechnique, Institut polytechnique de Paris, 91128 Palaiseau, France}
\address[address2]{Department of Civil Engineering, Stony Brook University, Stony Brook, NY 11794, USA}
\address[address3]{Division of Engineering and Applied Science, California Institute of Technology, Pasadena, CA 91125, USA}

\date{Pre-print: \today}

\begin{abstract}
The multi-scale nature of architectured materials raises the need for advanced experimental methods suitable for the identification of their effective properties, especially when their size is finite and they undergo extreme deformations. The present work demonstrates that state-of-the art image processing methods combined with numerical and analytical models provide a comprehensive quantitative description of these solids and their global behaviour, including the influence of the boundary conditions, of the manufacturing process, and of geometric and constitutive non-linearities. To this end, an adapted multi-scale digital image correlation analysis is used to track both elongations and rotations of particular features of the unit cell at the local and global (homogenized) scale of the material. This permits to observe with unprecedented clarity the strains for various unit cells in the structure and to detect global deformation patterns and heterogeneities of the homogenized strain distribution. This method is here demonstrated on elastic sheets undergoing extreme longitudinal and shear deformations. These experimental results are compared to non-linear finite element simulations, which are also used to evaluate the effects of manufacturing imperfections on the response. A skeletal representation of the architectured solid is then extracted from the experiments and used to create a purely-kinematic truss-hinge model that can accurately capture its behaviour. The analysis proposed in this work can be extended to guide the design of two-dimensional architectured solids featuring other regular, quasi-regular or graded patterns, and subjected to other types of loads.
\end{abstract}
\end{frontmatter}


\section{Introduction}
Architectured sheets are a particular class of two-dimensional solids whose patterned designs are tailored to achieve a variety of exceptional mechanical behaviours, including extreme extensibility, auxeticity and morphing capabilities \cite{Rafsanjani2016, Tang2017, Celli2018, Liu2018, Malomo2018, Choi2019, Boley2019, Guseinov2020, Liang2020}. They are increasingly seen  as applicable to fields ranging from stretchable electronics, medical and biomedical engineering \cite{Jiang2018a, Wang2019, Lee2019, Ali2013, Kapnisi2018}, to the sport equipment and textile industries \cite{Duncan2018, Foster2018, Wang2014b, Konakovic2016, Pattinson2019}, and they have witnessed significant advances in their design and fabrication.  When it comes to designing techniques, modern numerical methods such as shape and topology optimization \cite{Allaire2002a, Bendsoe2004} have become prevalent in this realm, leading to more sophisticated and often unimaginable geometries \cite{Wu2017a, Wang2017, Gao2018, Nika2019}. Present day techniques even permit to incorporate geometric non-linearity and manufacturability constraints in the design optimization \cite{Clausen2015, Wang2018, Zhang2019}.
At the same time, digitally controlled manufacturing techniques such as photo-lithography \cite{Tan2018}, 3-d printing \cite{Truby2016}, water jetting \cite{Coulais2018} and laser cutting \cite{Tang2017, Mizzi2020} now permit to fabricate architectured structures with unprecedented complexity and at a continuously decreasing cost.\smallskip

Despite these advancements, unleashing the potential of these systems demands advanced methods suitable for the experimental investigations on the deformation patterns and mechanical behaviour, which are to date in their early stages. In practice, specimens designed for mechanical characterization usually exhibit highly heterogeneous strain fields associated with: (i) their intrinsic multi-scale behaviour, that can be separated between the microscopic scale (material continuum) and the macroscopic scale (the global scale of the specimen); (ii) boundary layers that emerge from the boundary conditions and the finite size of the specimens; (iii) inherent anisotropic effective properties; (iv) sensitivity to shape imperfections. This high heterogeneity of the strain fields limits the level of identification that can be achieved from experimental measurements. For example, qualitative experimental insights on the behaviour of regions where macroscopic strains can be considered homogeneous have been reported in \cite{Mizzi2020}. As a consequence, only the central region of a specimen is typically used to validate numerical predictions \cite{Shan2015a, Clausen2015, Agnelli2020}, especially when one wants to compute the homogenised properties of the medium. It is to be noted that the interaction of scales is a key point for quantitatively understanding the behaviour of architectured solids. Experimentally, a precise separation between micro-scale and macro-scale kinematic fields, based on a first-order expansion of the fields, can be performed as illustrated in \cite{Rethore2015}. However, to the best of our knowledge, this technique has never been adopted in the context of architectured solids. The complexity of this interaction of scales has additionally motivated the development of reduced-order models \cite{Grima2005, Fu2016, Harkati2017}, to provide a better understanding of the underlying mechanisms and guidelines for design strategies. These reduced models often represent an idealised version of the unit-cell, and are inaccurate if not accompanied by robust experimental methods for the calibration of their parameters. On the opposite side of the spectrum of available numerical tools, lie models based on the complete description of the specimen, which are typically used to provide a direct term of comparison with experimental results \cite{Papka1999, Qiu2018, He2018}. The (often small) discrepancies between simulated and measured response have origins at multiple scales. They are either found at the microscopic scale, where the manufacturing process is a source of shape imperfections, or at the macroscopic scale, where applied boundary conditions may distort the unit cell pattern.\smallskip

The present work aims to demonstrate that various state-of-the art methods in image processing can be combined to provide comprehensive data on the multi-scale response of architectured sheets. Our procedure, applicable to any two-dimensional architectured solids, is here applied to investigate the deformation mechanisms of a soft auxetic sheet under extreme longitudinal and shear loading. The acquired images of the structure are first used to identify its exact geometry, which may differ from the designed one due imperfections in the fabrication process. Meshes are built directly from the identified shape and used both for the measurement of the full kinematic field (via Digital Image Correlation) as for the numerical computations (via the Finite Element Method). It is shown that doing so significantly improves the match between measurements and numerical predictions with respect to models that rely on the as-designed specimen geometry. This highlights the high sensitivity of the mechanical response of the specimen to geometrical imperfections. Then, we provide a two-scale analysis of the measured kinematic field: (i) at the continuum material level (microscopic scale) and (ii) at the unit cell level (macroscopic scale). This leads to the quantification of the macroscopic strain heterogeneities and the characteristic deformation patterns, which are influenced by the boundary conditions as well as the inherent Poisson's ratio of the micro-structure. The kinematic analysis is complemented by a procedure aiming at extracting the ``skeleton'' of the specimen from the experimentally-recorded images. This experimentally-extracted skeleton, whose shape changes during the deformation process, is then used to identify the parameters for an accurate reduced-order model of the architectured solid.\smallskip

The study is organized as follows: Section \ref{sec:material-method} provides details on fabrication, experimental setups, testing methods, material models and modelling strategy. The results are reported in Section \ref{sec:results-discussion}, and include the material constitutive law calibration, the multi-scale experimental analysis and the numerical simulations. The skeletal representation of the architectured sheet geometry is discussed in the same Section. A short summary in Section \ref{sec:conclusion} concludes the paper.


\section{Materials and methods} \label{sec:material-method}
\subsection{Fabrication of natural rubber architectured sheets}
\begin{figure}
\centering
\begin{minipage}{0.46\columnwidth}
\subf{\includegraphics[width=\columnwidth]{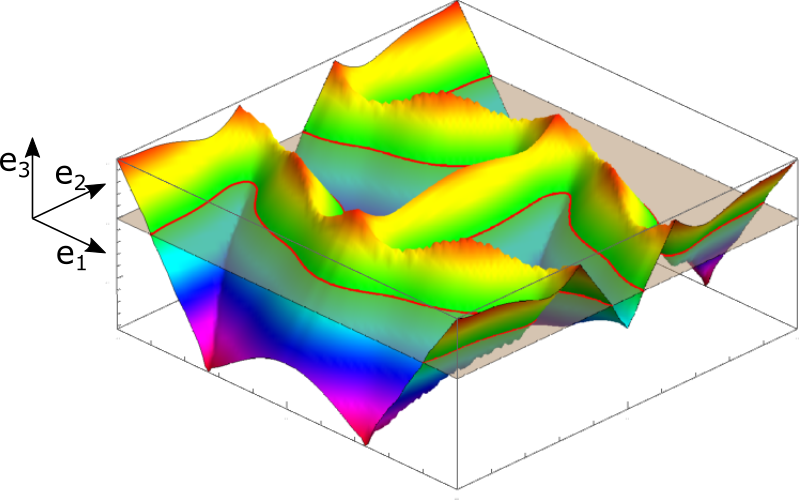}}{(a)}
\end{minipage}
\begin{minipage}{0.45\columnwidth}
\begin{equation*}
\begin{aligned}
&\text{material:}     & \phi(\vec{x}) > 0 \\
&\text{void:} & \phi(\vec{x}) < 0 \\
&\text{boundary:} & \phi(\vec{x}) = 0 \\
\end{aligned}	
\end{equation*}
\vfill
\end{minipage}
\subf{\includegraphics[width=\columnwidth]{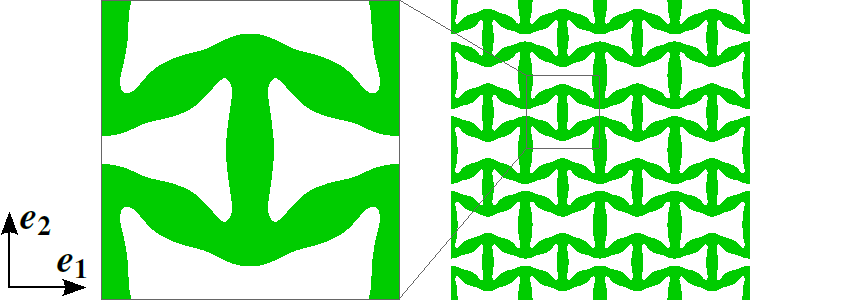}}
{(b) \hspace{3.1cm} (c)}
\caption{Geometry of the re-entrant honeycomb. (a) 3-d representation of the level set function (signed distance function) $\phi$, sliced by the plane $z = 0$. This function is obtained from a topology optimization procedure \protect \cite{Agnelli2020}. (b) Unit cell. (c) $4 \times 4$ repetitive array of unit cells.}
\label{fig:new_micro}
\end{figure}
To demonstrate our approach, we choose to analyse the periodic auxetic design recently proposed in \cite{Agnelli2020}. The design results from a topology optimization procedure combining the level set method and the asymptotic homogenization theory \cite{Allaire2004, Nika2019} aiming to minimize the auxetic behaviour \cite{Greaves2011}. The level set function $\phi$ serves as a base to define the material distribution in the unit cell (see \autoref{fig:new_micro}(a)), and is defined as the \textit{signed distance function}, for smoothness and regularity purposes.
Starting from the architecture provided in \cite{Agnelli2020}, we merely operate a vertical shift to obtain a symmetric design. The resulting unit cell is depicted in \autoref{fig:new_micro}(b,c). The designed geometry is a re-entrant honeycomb auxetic structure, with a couple of peculiar features. First, the structure is characterised by a repetitive alternation of two types of concave hexagons. Second, the trusses do not have constant width, \textit{i.e.} the linkages appear slightly thinner than the cores of the bars. This feature is similar to the bi-mode extremal material presented in \cite{Milton1995}.\smallskip

Mechanically, this architectured material carries an effective orthotropic behaviour (provided that the base material is isotropic \cite{Sanchez-Palencia1987}). Assuming an a-priori linear elastic behaviour implies that four coefficients need to be identified, namely the two effective Young's moduli, one effective Poisson's ratio and the effective shear modulus. A discussion on the elastic behaviour of the unit cell at small strain and on the identification of effective elastic coefficients is provided in \ref{appendix}.\smallskip

We fabricated three sorts of specimens consisting of periodic assemblages of the unit cell: two specimens designed for uniaxial tension along directions $\vec{e_1}$ and $\vec{e_2}$, hereafter referred to as specimens $T_1$ and $T_2$ respectively, and one specimen designed for a simple shear test, hereafter referred to as specimen $S$. The periodic array for each sample is set at:
\begin{itemize}[leftmargin =*]
\item $5 \times 8$ unit cells for the tensile specimen $T_1$ (see \autoref{fig:exp-setup}(b)),
\item $8 \times 5$ unit cells for the tensile specimen $T_2$,
\item a sequence of two lattices of $8 \times 5$ unit cells for the shear specimen $S$ (see \autoref{fig:exp-setup}(c), the arrangement is made to balance the torques).
\end{itemize}
For all specimens, the size of the square unit cell was set at $10\,\text{mm} \, \times \, 10 \,\mathrm{mm}$, yielding a $50 \, \mathrm{mm} \, \times \, 80 \,\mathrm{mm}$ lattice. The generated pattern is then completed by $50 \, \mathrm{mm} \, \times \, 10 \,\mathrm{mm}$ rectangular solid tabs that permits the clamping to the uniaxial testing machine. The specimens are laser cut from a $1.5\,\mathrm{mm}$-thick natural rubber sheet with a Universal ILS9 120 W laser cutter (single cut at $35\%$ power and $5\%$ speed). To avoid burning the rubber, the machine blows compressed air onto the part being cut. Prior to applying the speckle pattern on the specimens, these are thoroughly washed with standard dish-washing soap.

\subsection{Experimental setup and testing}
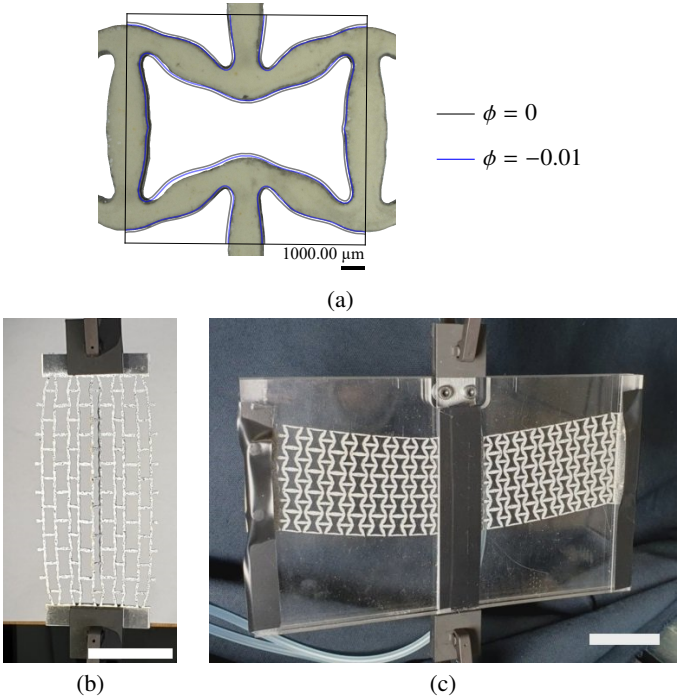
\begin{figure}
\centering
\subf{
\begin{tikzpicture}
\node at (0,0) {\includegraphics[height=0.4\columnwidth]{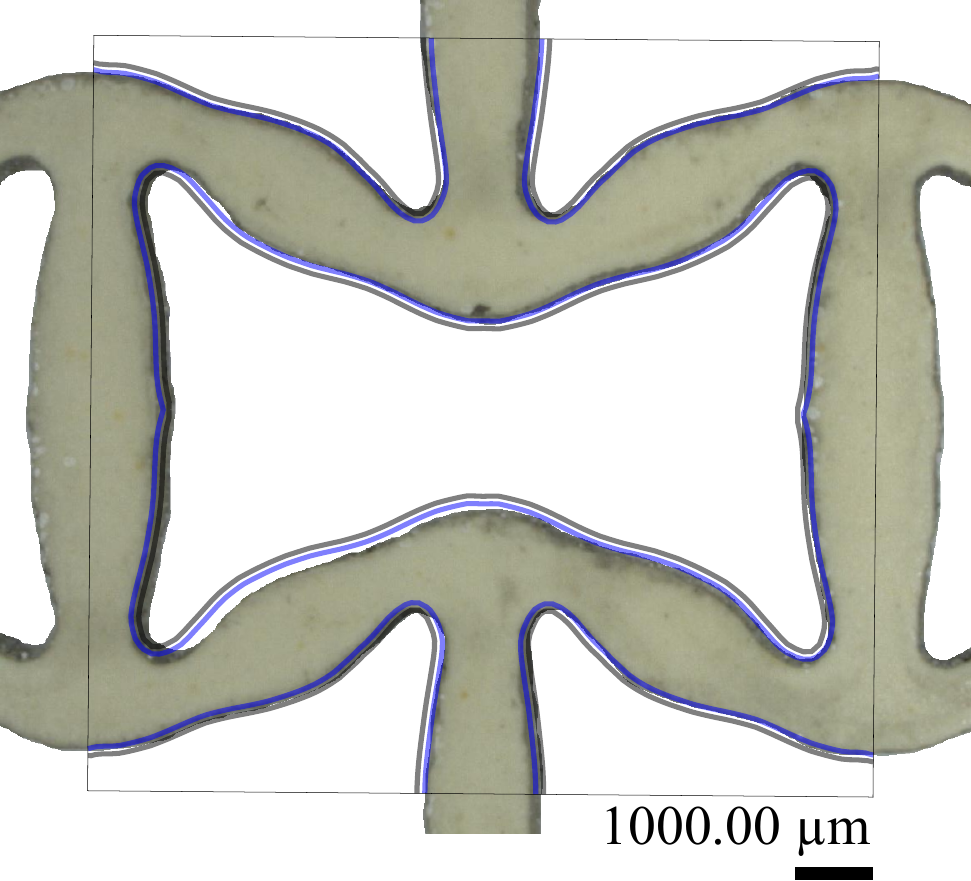}};
\node[right] (T) at (3, 0.3){$\phi = 0$};
\node[right] (B) at (3,-0.3){$\phi = -0.01$};
\draw[black] (2.5, 0.3)  -- (T);
\draw[blue]  (2.5,-0.3)  -- (B);

\end{tikzpicture}}{(a)}\\
\subf{\includegraphics[height=0.19\textheight]{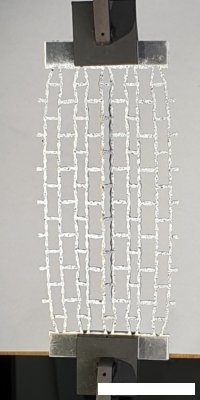}}{(b)}
\hfill 
\subf{\includegraphics[height=0.19\textheight]{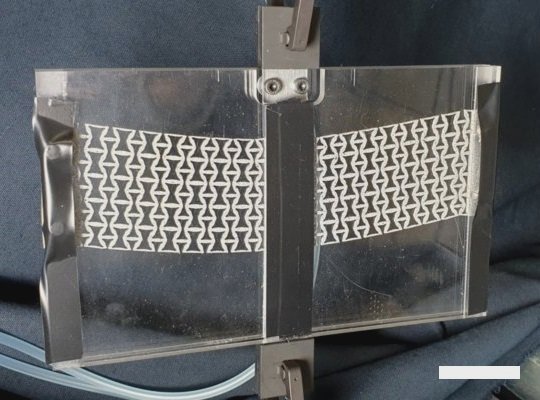}}{(c)}
\caption{(a) View of a unit cell of the fabricated specimen under a Keyence VHX-1000 optical microscope. (b-c) Setup for the tensile test (specimen $T_1$ here) and shear test (specimen $S$). In the shear test, PMMA confining plates are held together at their edges and are attached to the (sliding) upper grip. Conversely, the central rectangular rod is attached to the (fixed) lower grip. Scale bar is $40$ mm. }
\label{fig:exp-setup}
\end{figure}
To provide a complete characterization of this geometry, the evolving pattern transformations are investigated through uniaxial tensile and simple shear tests, as shown in \autoref{fig:exp-setup}(b,c). The experiments are conducted under displacement control at a quasi-static strain rate $\dot{\varepsilon} = 0.125 \,\mathrm{min}^{-1}$ up to $0.5$ effective engineering strain for the tensile test and up to $0.45$ effective engineering strain for the shear case. The tests are performed with an Instron $10\,\mathrm{kN}$ universal testing machine, with a mounted $50\,\mathrm{N}$ load cell with accuracy $\pm 0.1\,\mathrm{N}$. The specimens are clamped at both ends with metallic bars, to constrain their displacement (see \autoref{fig:exp-setup}(b)). The choice of hard clamp, which yield a strain heterogeneity in the specimens, was merely intended to facilitate the description of the boundary conditions in the numerical simulations. Recent works in literature \cite{Wang2019a} attempted to apply less constraining boundary conditions using rings and networks ensuring a homogeneous state of strain, at the cost of higher uncertainties on boundary conditions and stress state. For the shear test, a specific setup shown in \autoref{fig:exp-setup}(c) a specific setup is designed to arrange the specimen in the tensile machine. PMMA confining plates, preventing out-of-plane displacement, are held together at their edges and are attached to the (sliding) upper grip. Conversely, the central rectangular rod is attached to the (fixed) lower grip. The experiments were piloted using the Instron BlueHill software. Each mechanical test was recorded and used for full-field measurements by Digital Image Correlation (DIC). The recordings were obtained using a high-resolution digital camera (JAI Spark SP-20000-USB camera with a resolution of $5120 \times 3840$ pixels equipped with a Tokina AT-X Pro 100 mm F2.8 macro lens), mounted on a perpendicular axis with respect to the plane of the specimen. To improve the precision of the measurements, a gray scale speckle pattern was placed on the sample by aerosol spray. Using an in-built computer program, 8-bit gray scale sub-images were stored every second during the loading, with a resolution of $5064 \times 2438$ pixels for the tensile tests and resolution of $2292 \times 2488$ pixels for the shear test (the resolution for the shear is approximately two times smaller than in the tensile test because the camera was installed to record the whole specimen, yet only half of the specimen is useful for the observations).

\subsection{Local and global Digital Image Correlation} \label{sec:local_global_DIC}
All the results shown in this work make use of the the Digital Image Correlation technique (DIC) to extract the structure motion from acquired images during the test. DIC procedures are based on the comparison of subsequent pictures of the structure: given a \textit{reference} image $I_r$ and a \textit{current} image $I_i$, the problem consists in finding the displacement field $\vec u(\vec x)$ which minimizes the differences between the two images over a sub-domain $\Omega$:
\begin{equation}
\vec u(\vec x) = \underset{\vec{\xi}}{\mathrm{arg\,min}} \, 
\int \limits_\Omega{\big({ I_r[\vec x]-I_i[\vec x + \vec \xi(\vec x)] }\big)^2} \, d\Omega
\label{eq:DIC_problem}
\end{equation}
Given a parametrization of the the trial displacement field $\xi(x)$, this problem is usually solved using a Newton-Raphson procedure. The choice of this parametrization and the sub-domain $\Omega$ are the main elements that distinguish: (i) the \textit{local} approach, where $\Omega$ is restricted to a small image sub-domain over which the displacement is assumed to be homogeneous $\vec u(\vec x) = \vec a$ (thus sampling a uniform translation of $\Omega$) and (ii) the \textit{global} approach where the displacement is defined over a finite-elements mesh covering the full domain of interest $\Omega$ (i.e $\vec u(\vec x) = \vec N(\vec x) \cdot \vec a$ with $\vec N(\vec x)$ containing the finite element shape functions).\smallskip

While the comparison between both approaches in terms of efficiency and accuracy is still a hot topic in the community \cite{Hild2012, Rethore2017}, they are both used for different purposes in the present study. Indeed, the global approach assumes the displacement field continuity over the domain $\Omega$, which is well suited for the study of the structure at the \textit{microscopic} scale (corresponding to the material continuum). Conversely, the local approach is employed to follow the motion of isolated points at the \textit{macroscopic} scale (corresponding to the pattern periodicity), for example to study the motion of the corner nodes of each unit cell.\smallskip
	
All DIC results presented in this paper are obtained from an in-house academic code written by means of MATLAB scripts. For the global approach, simplex $P1$ triangular elements are used. The meshes are generated using the \texttt{DistMesh} procedure proposed by Persson \cite{Persson2004} with the following steps: first, a binary mask is obtained from the reference image (where the specimen is unstrained). Second, a distance transform is applied on the mask to obtain the \textit{experimental} level-set function sampling the specimen boundaries. Finally, the \texttt{DistMesh} procedure is applied with the obtained level-set function as input. We chose an edge length of $10$ pixels, sufficient to capture the localization of strains in the structure while keeping a good DIC resolution (sub-pixel accuracy). Hereafter, the resulting mesh is referred as to $\mathcal{M}_{i}^{DIC}$ ($i$ denotes the specimen name).

\subsection{Numerical Simulations}
\paragraph{Finite element method implementation} \label{sec:finite_element_model}
Finite element computations are undertaken under the assumption of large strains plane stress using the finite element solver Cast3M 2020 ({\tt www-cast3m.cea.fr}). In the simulations, the conditions of the mechanical tests are exactly reproduced, \textit{e.g.} the sample is loaded in with a prescribed displacement at the two ends. In both cases, the specimen is meshed with $P2$ triangle elements. The geometry of the specimen used for the computations is obtained following two strategies:
\begin{itemize}[leftmargin =*]
\item from the \textit{theoretical} level set function $\phi$, using image processing to detect and extract the 0-level contour image of the level set function. Hereafter this mesh is referred as to $\mathcal{M}_{i}^{\phi \, = \,0}$ ($i$ denotes the specimen name). For all specimens, the total numbers of elements and nodes are 80,000 and 171,534, respectively. $\mathcal{M}^{\phi \, = \,0}$ is perfectly periodic, \textit{i.e.} it does not embed any geometrical defects;
\item from the \textit{experimental} mesh $\mathcal{M}_i^{DIC}$ (used for the global DIC presented in section \ref{sec:local_global_DIC}). The total numbers of elements and nodes for the FE model are 78,380 and 166,982, respectively. By comparison to the theoretical mesh $\mathcal{M}_i^{\phi \, = \,0}$, $\mathcal{M}_i^{DIC}$ captures several geometrical imperfections induced by the fabrication process and by the positioning of the specimen in the tensile machine.
\end{itemize}

\paragraph{Rubber material models} The constitutive behaviour of natural rubber is modelled as an incompressible hyperelastic material. Let $\vec{F} = \frac{\partial \vec{x}}{\partial \vec{X}}$ denote the deformation gradient mapping a material point from the reference position $\vec{X}$ to its current location $\vec{x}$. We adopt the Mooney-Rivlin model \cite{Mooney1940,Rivlin1948}, which is normally acceptable for intermediate elongations, \textit{i.e.} between 50  -100\%. The strain energy function of Mooney-Rivlin hyperelastic constitutive law is expressed as a function of strain invariants $I_1,\, I_2,\, I_3=J^2$ of the left Cauchy-Green tensor $\vec{B} = \vec{F} \vec{F}^T$. The strain energy density function takes the form:
\begin{equation}
W= C_{10} (I_1 - 3) + C_{01} (I_2 - 3) + \frac{1}{d} (J - 1)^2
\end{equation}
where $C_{10}, \, C_{01}$ and $d$ are material parameters. For the case of an incompressible Mooney-Rivlin material under uniaxial elongation, $\lambda_1 = \lambda \,$ and $\lambda_2 = \lambda_3 = 1/{\sqrt {\lambda}}$. Then the true stress (Cauchy stress) differences can be calculated as: 
\begin{equation}
\begin{aligned}
\sigma_{11} - \sigma_{33} &= 2 C_{10} ( \lambda^2 - \frac{1}{\lambda}) - 2 C_{01} ( \frac{1}{\lambda^2} - \lambda^2) \\
\sigma_{22} - \sigma_{33} &= 0\\
\end{aligned}
\end{equation}
In the case of simple tension, $\sigma _{22}=\sigma _{33}=0$. Then we can write:
\begin{equation}
\sigma _{11} = \left(2 C_{10} + {\cfrac {2C_{01}}{\lambda }}\right) 
\left(\lambda ^{2}-{\cfrac {1}{\lambda }}\right)
\end{equation}
and the engineering stress (force per unit reference area) for an incompressible Mooney–Rivlin material under simple tension can be calculated using $\sigma_{11}^{\mathrm{eng}} = \sigma_{11} \lambda_2 \lambda_3 = \sigma_{11}/ \lambda = \sigma_{11}/ (1 + e_{11}^{\mathrm{eng}})$. Hence:
\begin{equation}
\begin{aligned}
\sigma_{{11}}^{{{\mathrm {eng}}}}=\left(2C_{10}+{\frac {2C_{01}}{\lambda }}\right)\left(\lambda -\lambda ^{{-2}}\right)\\
\sigma_{{11}}^{{{\mathrm {eng}}}}=\left(2C_{10}+{\frac {2C_{01}}{1 + e_{11}^{\mathrm{eng}}}}\right)\left(1 + e_{11}^{\mathrm{eng}} - \frac{1}{(1 + e_{11}^{\mathrm{eng}})^2}\right)
\end{aligned}
\end{equation}


\section{Results and discussion}\label{sec:results-discussion}
\subsection{Numerical simulations}
\begin{figure}
\centering
\subf{\includegraphics[width=0.4\columnwidth]{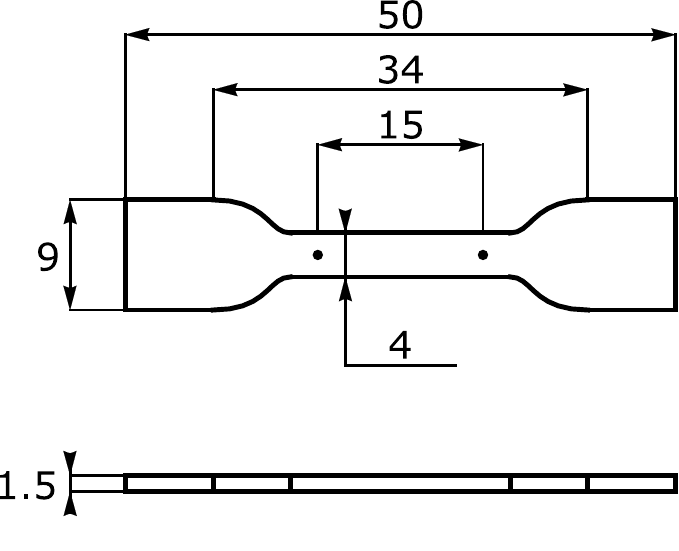}}{(a)}
\hfill
\subf{\includegraphics[width=0.49\columnwidth]{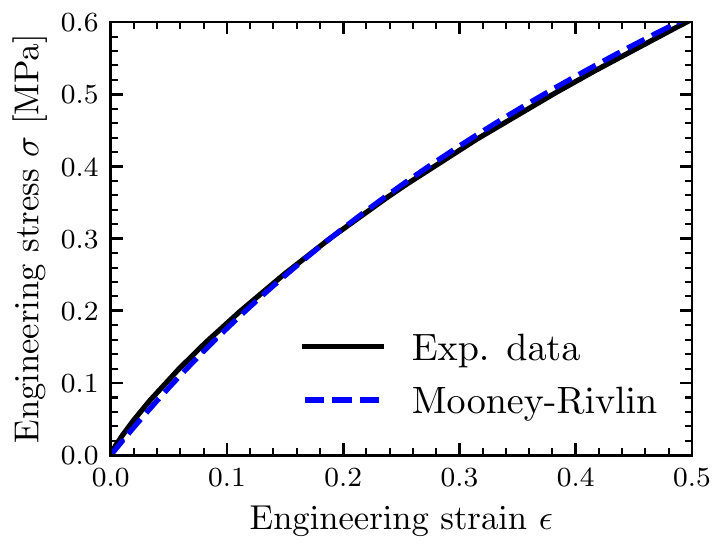}}{(b)}
\caption{(a) The dogbone geometry with its dimensions in mm. (b) Measured engineering stress-strain response under uniaxial tension. The Mooney-Rivlin hyperelastic model is employed to fit the stress–strain response and calibrate material parameters.}
\label{fig:uniaxial_rubber}
\end{figure}
\paragraph{Calibration of material parameters} The mechanical behaviour of natural rubber is identified from uniaxial tensile tests. Dogbone specimens are fabricated using a cutting die to make specimens for uniaxial tension (the dimensions of test specimens are depicted on \autoref{fig:uniaxial_rubber}(a)) and are subjected to the uniaxial tensile tests with a speed of $10\,\mathrm{mm/min}$. The measured engineering stress–strain response is shown in \autoref{fig:uniaxial_rubber}(b). It is shown that the Mooney-Rivlin model is suitable to capture the tensile behaviour well up to 0.5 engineering strain for this natural rubber. The material coefficients $C_{10} = 0.199169\,\mathrm{MPa}$ and $C_{01} = 0.134212\,\mathrm{MPa}$ in the Mooney-Rivlin model for this natural rubber are identified by a non-linear fit from the experimental data
\begin{figure*}
\centering
\subf{\includegraphics[height=0.32\textwidth]{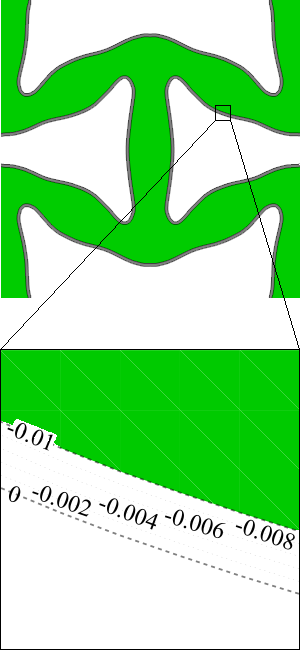}}{(a) Cut levels for $\phi$}
\hfill
\subf{\includegraphics[height=0.32\textwidth]{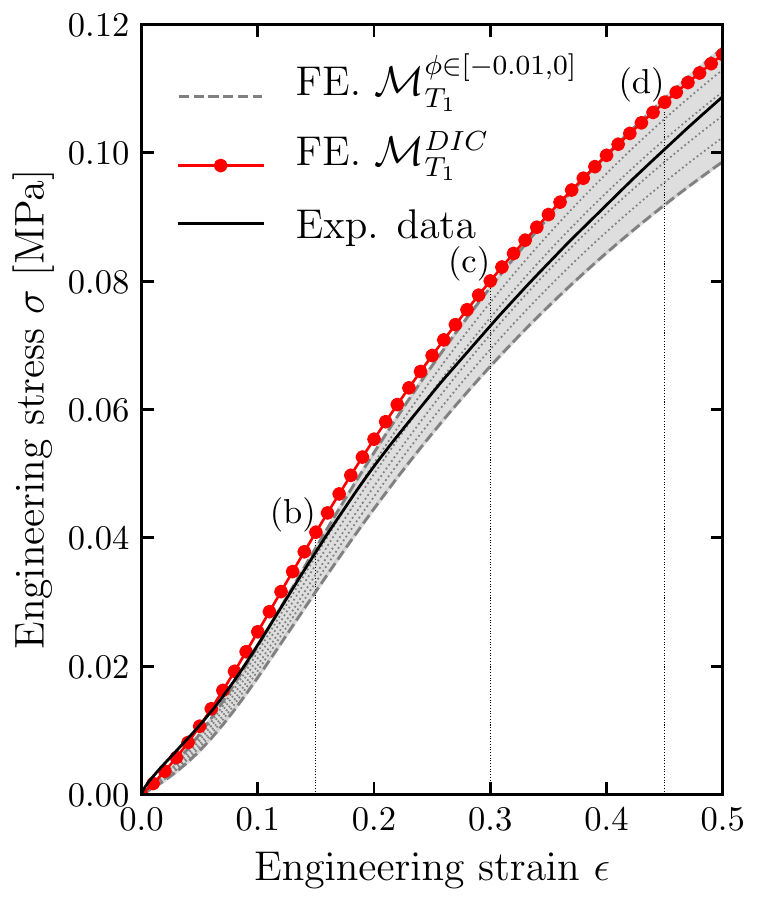}}{(b) $T_1$}
\subf{\includegraphics[height=0.32\textwidth]{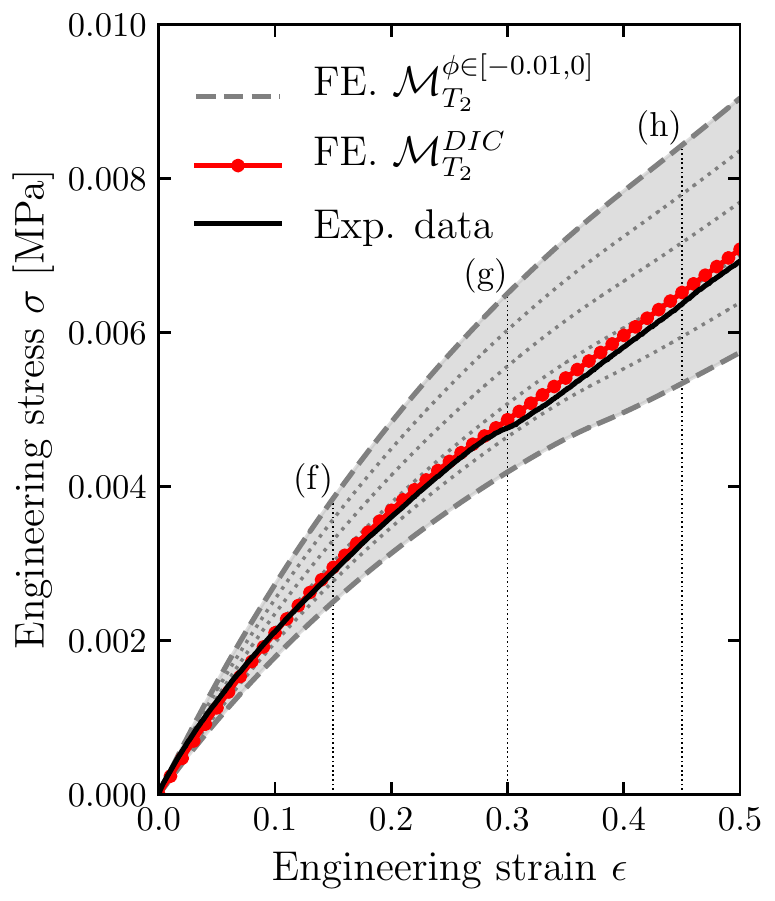}}{(c) $T_2$}\subf{\includegraphics[height=0.32\textwidth]{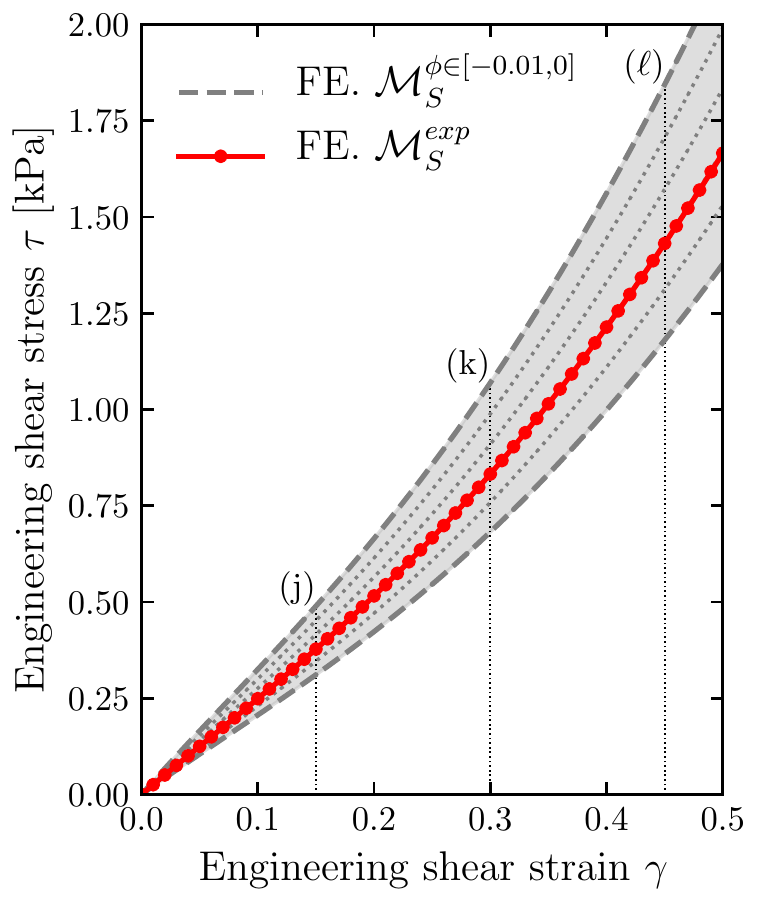}}{(d) $S$}
\caption{(a) Unit cell contour defined by the level set function $\phi$ with varying cutting heights. (b-d) Effective stress-strain curves for the structure. Comparison between experiments and numerical simulations with the Mooney-Rivlin hyperelastic model. The shaded gray areas encompass the stress-strain curves for $\phi \in [-0.1, 0]$. The letters appearing at 0.15, 0.3 and 0.45 effective strains refer to the deformed shapes in \autoref{fig:uniaxial_honeycomb}.}
\label{fig:stress-strain_honeycomb}
\end{figure*}
\paragraph{Shape sensitivity analysis} We first report the measured engineering stress-strain curves for all tests (see \autoref{fig:stress-strain_honeycomb}). For tensile tests (specimens $T_1$ and $T_2$), the experiments are juxtaposed to the numerical results (for the shear, the frictions in the setup do not allow to obtain an experimental estimate of the load.) \autoref{fig:stress-strain_honeycomb}(b) and even more \autoref{fig:stress-strain_honeycomb}(c-d) reveal a significant gap in stiffness between the numerical predictions on the theoretical mesh $\mathcal{M}^{\phi=0}$ (stiffest dashed gray curve) and on the experimental mesh $\mathcal{M}^{DIC}$. The latter model is in better agreement with the experiments (black curves). The strong differences between the two approaches in the numerical analyses suggest that the material effective stiffness is highly sensitive to the shape uncertainties induced by the laser cutting. To analyse the sensitivity of the mechanical behaviour to shape uncertainty, additional numerical simulations are carried out using \textit{eroded} theoretical meshes, \textit{i.e.} by progressively reducing the size of the trusses. In practice, we operate an erosion of the contour by introducing a negative offset to the signed distance function $\phi$ of \autoref{fig:new_micro}(a). The behaviour for offsets varying between $-0.1$ and $0.$ with a step of $0.02$ is shown in \autoref{fig:stress-strain_honeycomb}(a). The experimental stress-strain curves of specimen $T_2$ (\autoref{fig:stress-strain_honeycomb}(c)) are most similar to the eroded model with the level set shifted by $-0.06$. Using the properties of the signed distance function $\phi$, the experimental specimen is expected to be fabricated with trusses that are roughly $120 \, \mathrm{\mu m}$ thinner than expected. This gap to the laser cutting process. In hindsight, observing the specimens under an optical microscope (see \autoref{fig:exp-setup}(a)) confirms that these are thinner than expected and also reveals that the error on the thickness is not constant along the trusses. In the following, the simulations performed on the experimental mesh $\mathcal{M}^{DIC}$ are used for the comparison with experiments and general validation.

\begin{figure*}[ht]
\centering
\begin{tabularx}{\textwidth}{cXXX}
\subf{\includegraphics[width=0.128\textwidth]{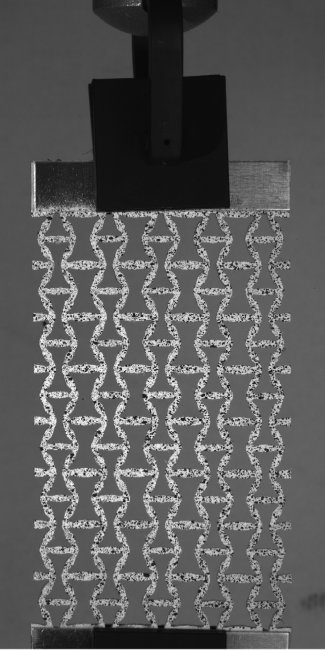}}{(a) $\varepsilon = 0\%$} &
\subf{\includegraphics[width=0.128\textwidth]{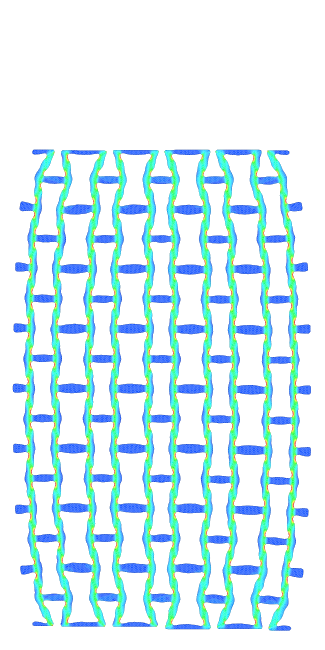}}{(b) $\varepsilon = 15\%$}
\subf{\includegraphics[width=0.128\textwidth]{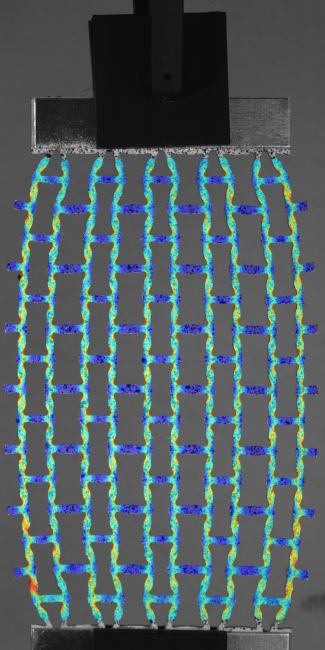}\\
\includegraphics[scale=0.9]{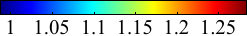}}{} &
\subf{\includegraphics[width=0.128\textwidth]{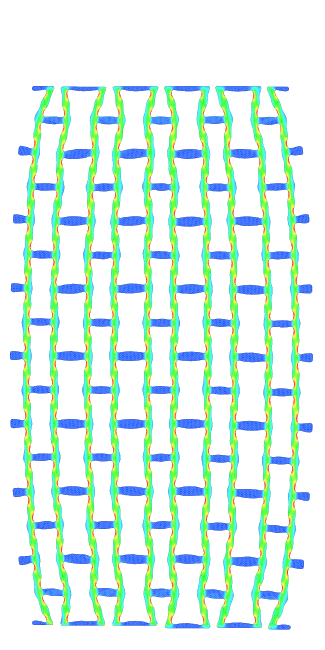}}{(c) $\varepsilon = 30\%$}
\subf{\includegraphics[width=0.128\textwidth]{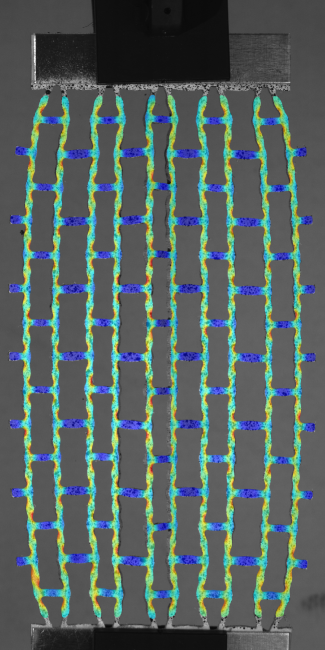}\\
\includegraphics[scale=0.9]{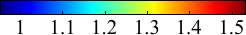}}{}&
\subf{\includegraphics[width=0.128\textwidth]{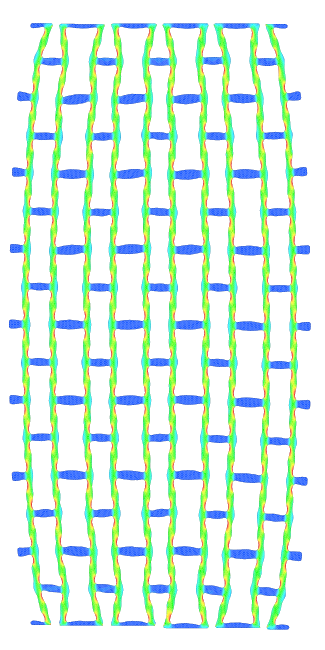}}{(d) $\varepsilon = 45\%$}
\subf{\includegraphics[width=0.128\textwidth]{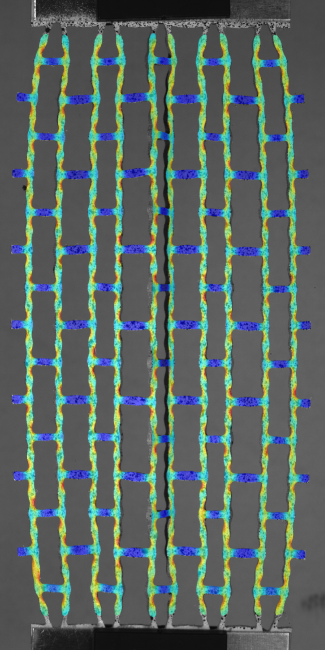}\\
\includegraphics[scale=0.9]{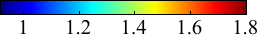}}{\\} \\
\subf{\includegraphics[width=0.128\textwidth]{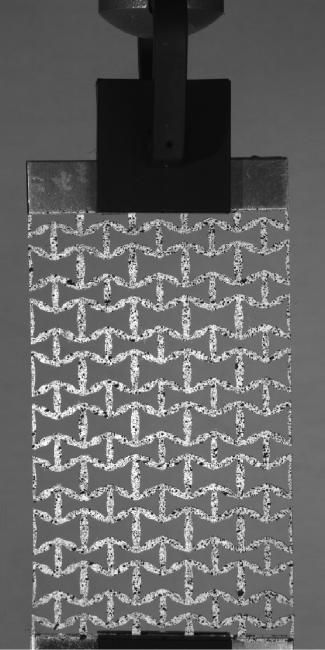}}{(e) $\varepsilon = 0\%$} &
\subf{\includegraphics[width=0.128\textwidth]{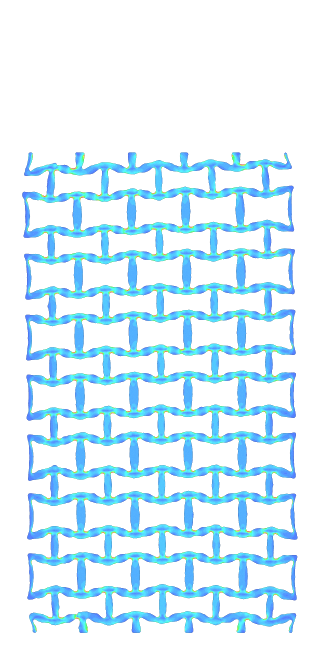}}{(f) $\varepsilon = 15\%$}
\subf{\includegraphics[width=0.128\textwidth]{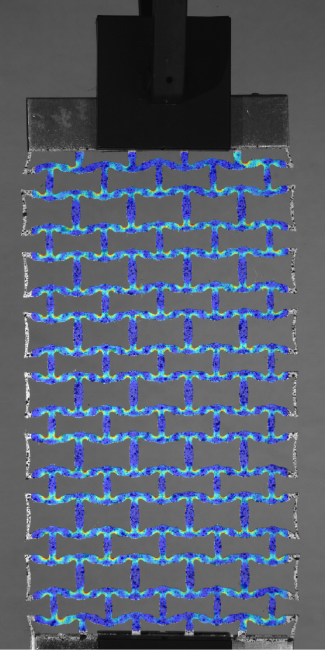}\\
\includegraphics[scale=0.9]{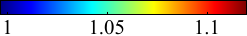}}{}&
\subf{\includegraphics[width=0.128\textwidth]{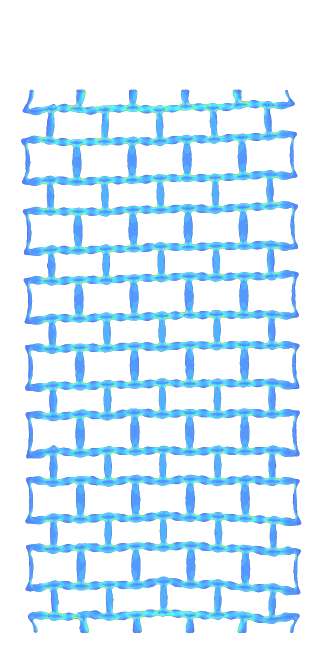}}{(g) $\varepsilon = 30\%$}
\subf{\includegraphics[width=0.128\textwidth]{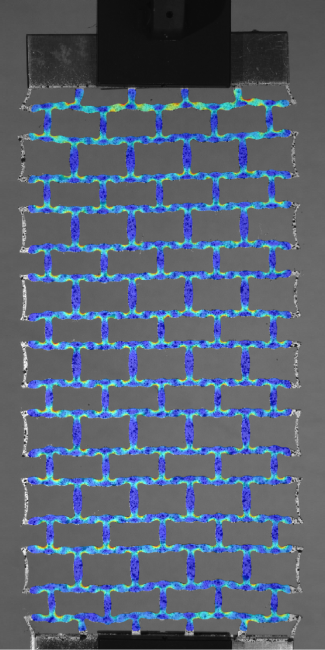}\\
\includegraphics[scale=0.9]{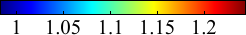}}{}&
\subf{\includegraphics[width=0.128\textwidth]{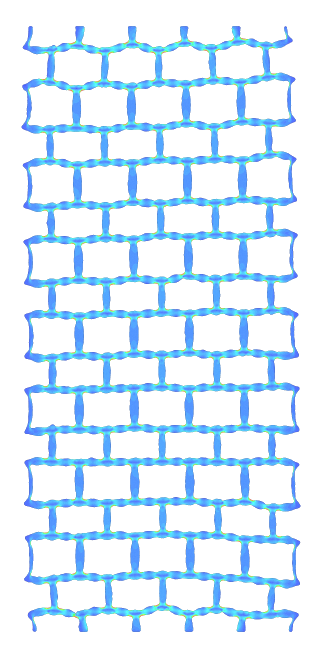}}{(h) $\varepsilon = 45\%$}
\subf{\includegraphics[width=0.128\textwidth]{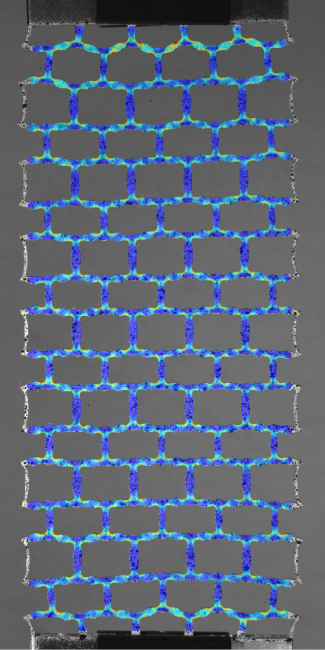}\\
\includegraphics[scale=0.9]{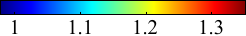}}{\\} \\
\subf{\includegraphics[width=0.128\textwidth]{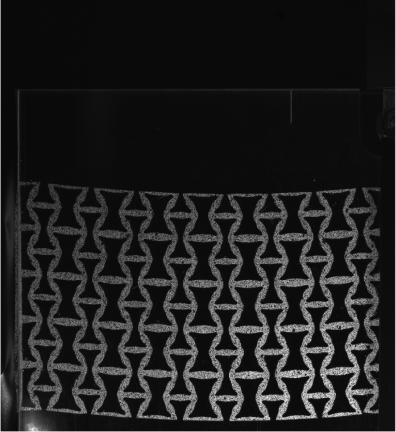}}{(i) $\gamma = 0\%$}& 
\subf{\includegraphics[width=0.128\textwidth]{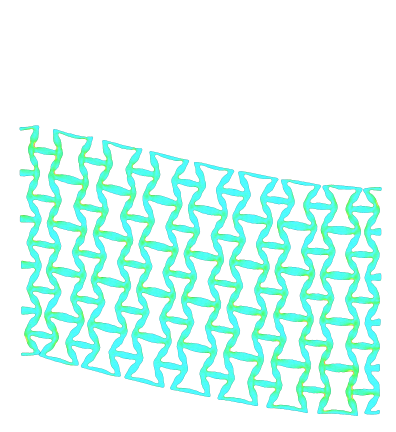}}{(j) $\gamma = 15\%$}
\subf{\includegraphics[width=0.128\textwidth]{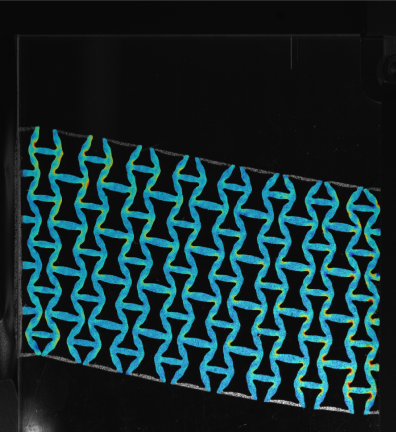}\\
\includegraphics[scale=0.9]{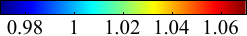}}{}&
\subf{\includegraphics[width=0.128\textwidth]{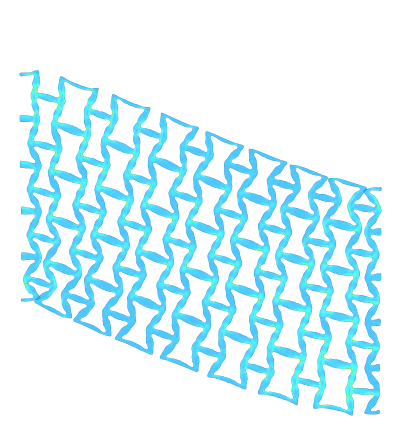}}{(k) $\gamma = 30\%$}
\subf{\includegraphics[width=0.128\textwidth]{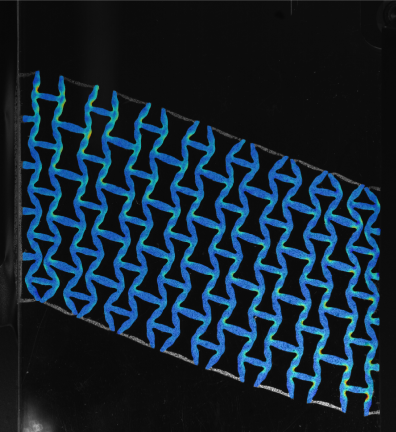}\\
\includegraphics[scale=0.9]{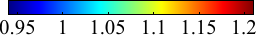}}{}&
\subf{\includegraphics[width=0.128\textwidth]{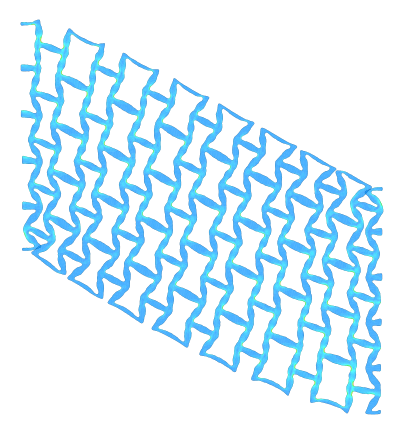}}{($\ell$) $\gamma = 45\%$}
\subf{\includegraphics[width=0.128\textwidth]{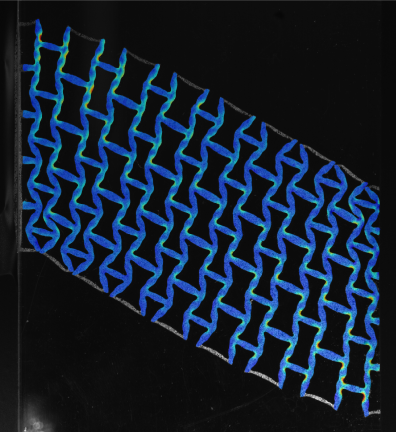}\\
\includegraphics[scale=0.9]{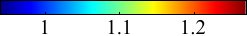}}{}
\end{tabularx}
\caption{Numerical and experimental deformed configurations of specimens $T_1$ (a-d), $T_2$ (e-h) and $S$ (i-$\ell$) at different levels of imposed engineering effective strain: $0, 0.15, 0.30$ and $0.45$. The principal stretch $\lambda_1$ is plotted as a colormap in each figure. The colorbar is the same for both the numerical and experimental results.}
\label{fig:uniaxial_honeycomb}
\end{figure*}

\subsection{Two-scale kinematic analysis}
\paragraph{Scale of the sheet material continuum} For all the tests, the acquired images of the structure are reported in \autoref{fig:uniaxial_honeycomb} for stages corresponding to $0$, $0.15$, $0.3$ and $0.45$ engineering strain. The principal stretch field $\lambda_1$ resulting from the global-DIC procedure performed on a full set of acquired pictures is superimposed to the images. Following the procedure described in section \ref{sec:local_global_DIC}, the \textit{experimental} mesh $\mathcal{M}^{DIC}$ used to perform the DIC is defined at the reference stage. The obtained displacements fields permit a further comparison with predictions and give an insight on the deformation mechanism of the samples, \textit{i.e.} how the structure moves and deforms.\smallskip

In all the tests, the distribution of the elongation (\autoref{fig:uniaxial_honeycomb}) obtained from the displacement field in both full-field measurement indicates that the strain field is mostly concentrated on the hinges of the structure. This emphasizes the predominance of structural deformation at small strain, where different parts of the \textit{lattice} behave as rigid struts and deformable hinges, in spite of the soft natural rubber. For the tensile tests, a lateral expansion indicating a negative Poisson's ratio is visible in both $T_1$ and $T_2$ specimens. Despite these general observations, some discrepancies can be noticed between the two tensile specimens. First, the amount of transverse strain is obviously different between specimen $T_1$ and $T_2$, expressing the orthotropic nature of the design. Second, while the most of strain is localised at the hinge regions in the specimens $T_2$ and $S$, a clear elongation of the members is identified on specimen $T_1$.\smallskip

\autoref{fig:uniaxial_honeycomb}(b-d) shows that specimen $T_1$ undergoes a positive strain in the trusses under tension (at 0.15 effective strain, $\lambda_1 \approx 1.15$ in green), whereas the perpendicular members exhibit negative strain (with $\lambda_1 < 1$). This transverse compressive state is responsible for an out-of-plane buckling at $\sim 0.15$ effective engineering strain. Beyond this stage, a wrinkling deformation is observed \textit{i.e.} each transverse branch becomes corrugated (see the central unit cells in \autoref{fig:exp-setup}(b), \autoref{fig:uniaxial_honeycomb}(c-d) and \href{run:Movie1_T1_OneCell_GL_Strains.mp4}{Movie 1}). This particular instability is typical of the clamped boundary conditions imposed on the specimen, responsible for compressive stresses that develop in the transverse direction \cite{Cerda2002}. The buckling and post-buckling modelling, beyond the scope of the paper, is neither accounted nor permitted in the two-dimensional finite element model. Since DIC measurement is also based on a 2-d model, the out of plane deformation appears as compression state in the stretch field in \autoref{fig:uniaxial_honeycomb}(c,d). Looking at \autoref{fig:stress-strain_honeycomb}(b), this illustrates why the numerical simulation (curve in red) perfectly matches the experiment (curve in black) until $0.15$ effective engineering strain, while it tends to overestimate the effective stress at larger strains. The maximal relative error between the experiment and the simulation is of $9.5\%$.\smallskip

Specimen $T_2$ remains mostly unstrained at the core of the trusses throughout the test ($\lambda_1 \approx 1.$ in blue). The specimen remained in the plane during the whole test. However, unit cells located at its edges  experienced snap-through instabilities just before 0.3 effective engineering strain. Indeed, the buckled cells that were almost unstrained in \autoref{fig:uniaxial_honeycomb}(f) become the most strained in \autoref{fig:uniaxial_honeycomb}(g,h). The full movie of the tensile test provided in the supplementary material permits to better appreciate the effect (see \href{run:Movie2_T2_Lambda1_Full.mp4}{Movie 2}). This effect is observed in both the experiments and the numerical simulations. This feature is also detected in \autoref{fig:stress-strain_honeycomb}(b) where a local change in the slope of the stress-strain curve corresponding to the relaxation of the center cells accompanying the edge cells snap-through is identified. Note that the samples are \textit{monostable} unlike the examples of \cite{Rafsanjani2016}, \textit{i.e.} once unloaded, the specimens return to their initial configurations. In \autoref{fig:stress-strain_honeycomb}(c), the numerical simulation (curve in red) correctly matches the experiment (curve in black) until $0.5$ effective engineering strain. The small gap that appears around $0.3$ effective engineering strain is attributed to the snapping effect which is not captured the numerical stress-strain curve. The maximal relative error between the experiment and the simulation is of $3\%$.\smallskip

Regarding the shear test, the $S$ specimen is mounted horizontally (refer to \autoref{fig:exp-setup}(b)). Therefore, its own weight induces an initial bending visible in \autoref{fig:uniaxial_honeycomb}(i). Nonetheless, the role of the weight rapidly becomes negligible as the applied shear load increases ($\gamma > 0.1$). As we establish a relative good agreement between simulation and experiments under uniaxial tension (besides structural instabilities that were not accounted), the finite element method permits to estimate the stress distribution during shear test (see \autoref{fig:stress-strain_honeycomb}(c)). Regardless of the shear set-up, we remark that the values of the load (maximal effective stress expected of $1.75 \, \mathrm{kPa}$, yielding a resultant load of $0.14$ N) would have been too small to be precisely measured with our experimental tools.
%
\begin{figure*}
\def\figHeight{0.27\textwidth}
\def\figWidth{1.14*\figHeight}
\def\SfigHeight{0.173\textwidth}
\centering
\begin{tabularx}{\textwidth}{cXXX}
\subf{
\begin{tikzpicture}[baseline]
    \begin{axis}[%
    	scale only axis,
        width=\figWidth, height=\figHeight, at={(0,0)},
		xlabel={Longitudinal strain $\varepsilon_{11}$},
		ylabel={Transverse strain $\varepsilon_{22}$},
		xlabel near ticks, ylabel near ticks,
        grid=both,
        grid style={dotted, line width=.1pt, draw=black!50},
        enlargelimits=false,
        axis on top,
        clip=false
            ]
        \addplot[] 
        	graphics[xmin=0,ymin=-0.05,xmax=0.45,ymax=0.30]{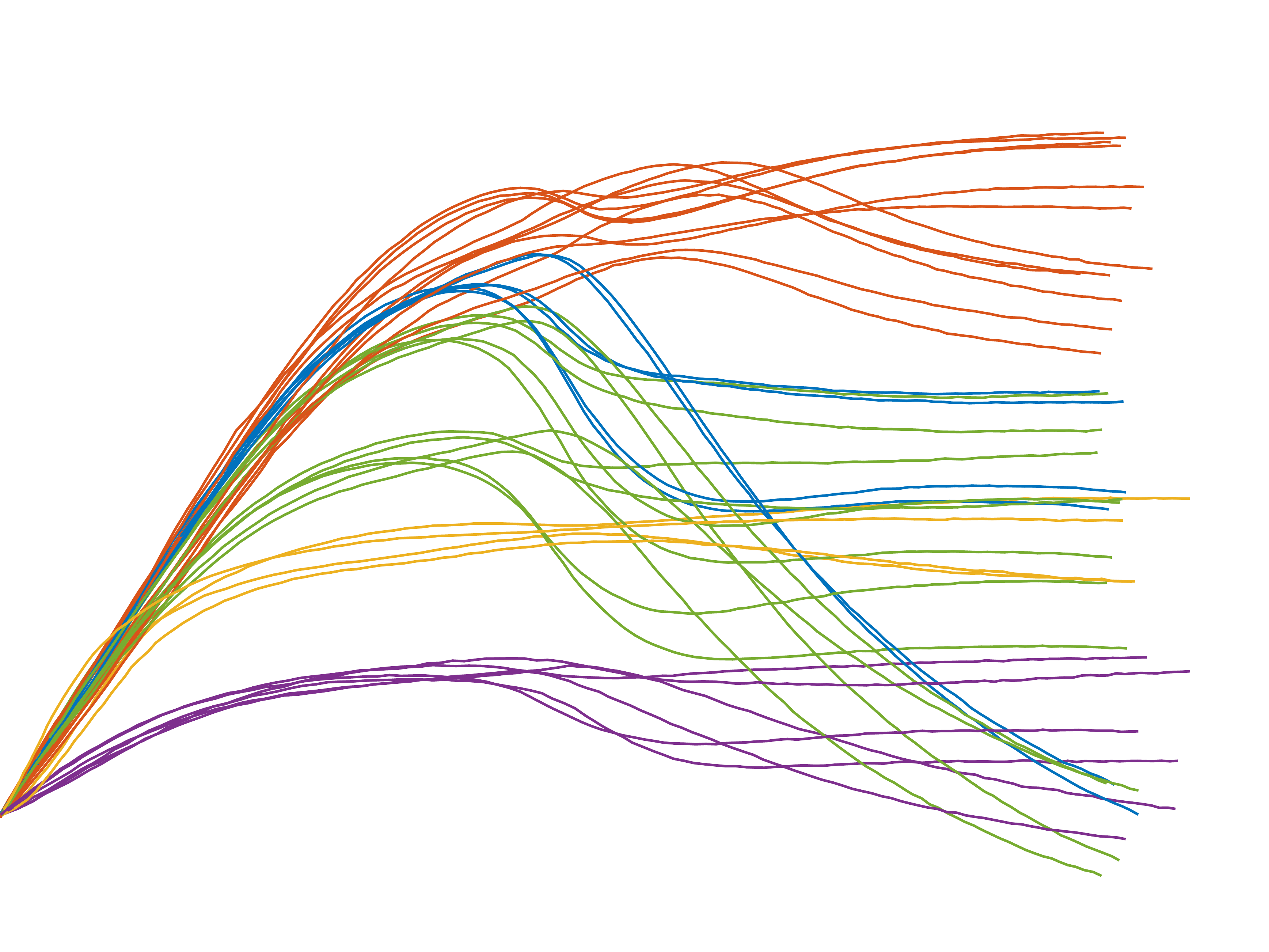};
        \draw[dashdotted] (0,0) -- (0.167,0.25) ;
		\node[anchor=west] at (0.167,0.275) {$\nu = -1.5$};	
        \node[below right] at (rel axis cs: 0.01,0.99) 
        	{\includegraphics[scale = 0.25]{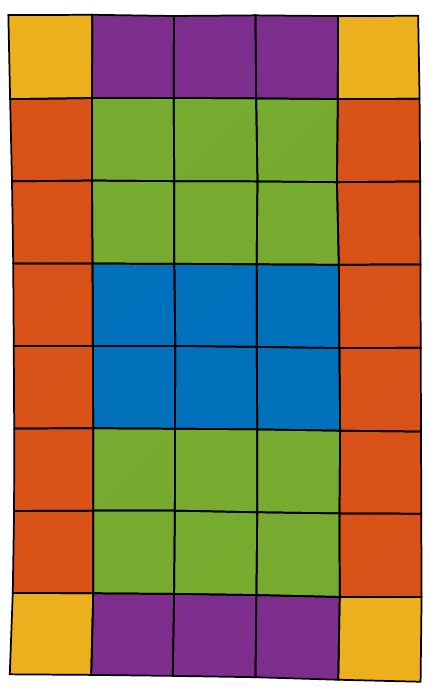}};       
    \end{axis}
\end{tikzpicture}}{(a) $T_1$}&
\subf{\includegraphics[height=\figHeight]{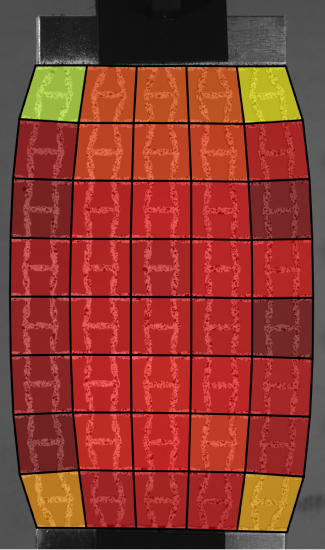}\\
\includegraphics{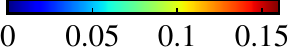}}{(b) $\varepsilon_{11}$}&
\subf{\includegraphics[height=\figHeight]{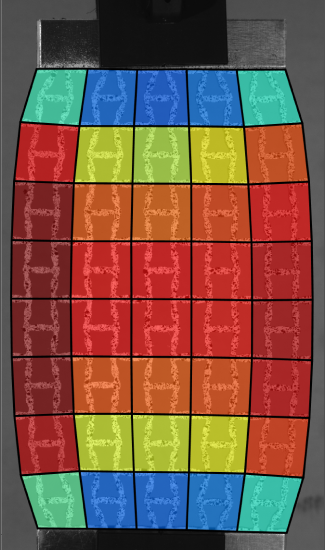}\\
\includegraphics{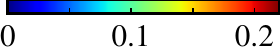}}{(c) $\varepsilon_{22}$}&
\subf{\includegraphics[height=\figHeight]{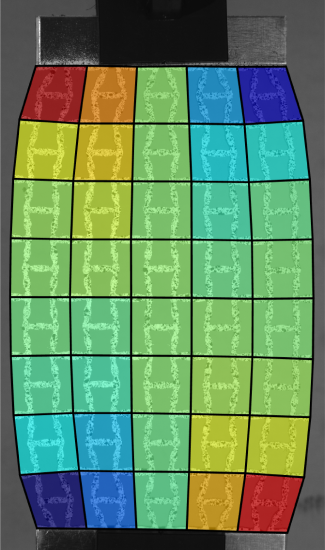}\\
\includegraphics{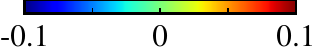}}{(d) $\varepsilon_{12}$}\\\\
\subf{
\begin{tikzpicture}[baseline]
    \begin{axis}[%
		scale only axis,
		width=\figWidth, height=\figHeight, at={(0,0)},
		xlabel={Longitudinal strain $\varepsilon_{11}$},
		ylabel={Transverse strain $\varepsilon_{22}$},
		xlabel near ticks, ylabel near ticks,
		yticklabel style={/pgf/number format/fixed},
        grid=both,
        grid style={dotted, line width=.1pt, draw=black!50},
        enlargelimits=false,
        axis on top,
        clip=false
            ]
        \addplot[]
        	graphics[xmin=0,xmax=0.5,ymin=-0.02,ymax=0.10]{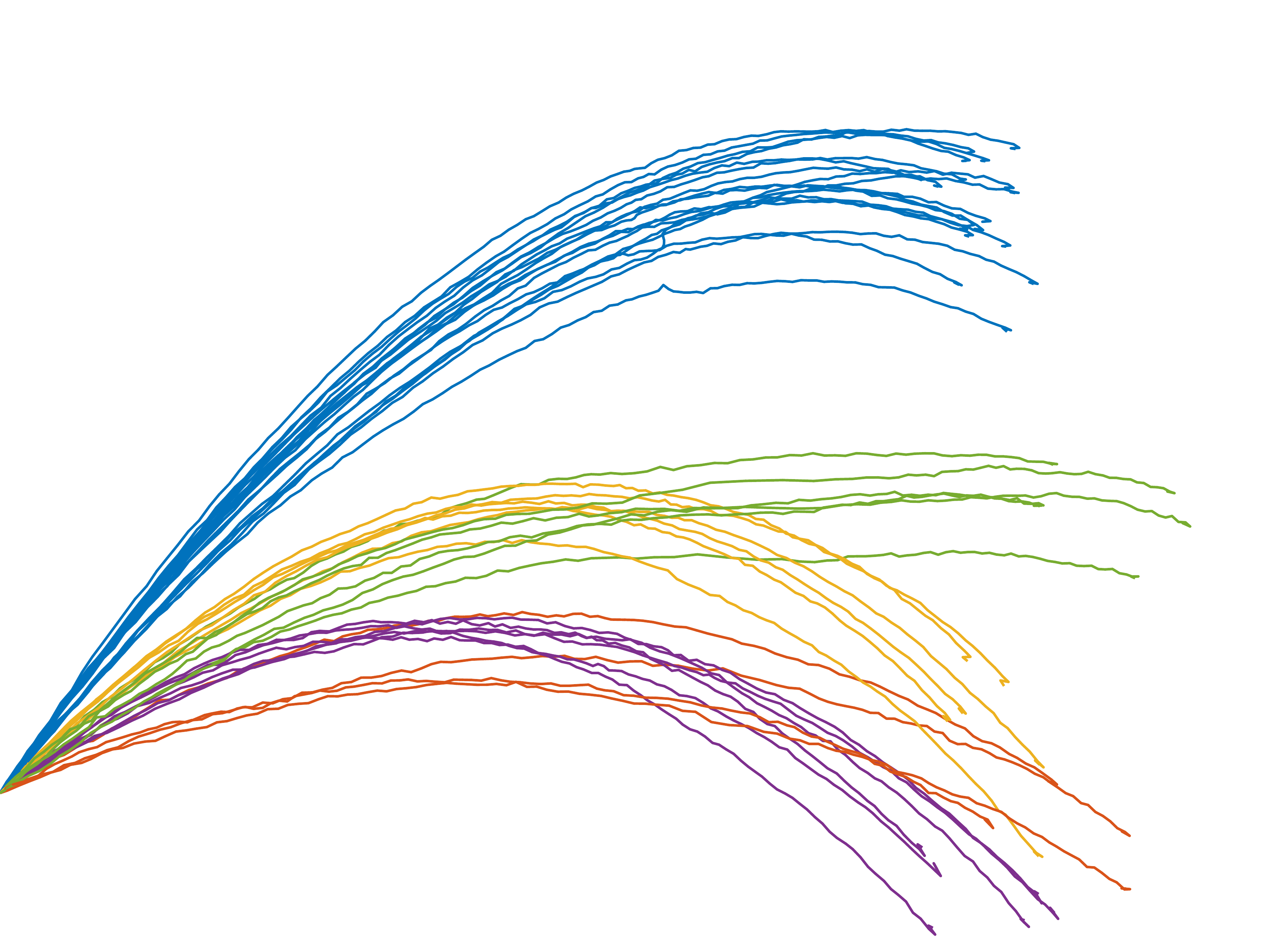};
        \draw[dashdotted] (0,0) -- (0.225,0.09) ;
		\node[anchor=west] at (0.225,0.092) {$\nu = -0.4$};		        
        \node[below right] at (rel axis cs: 0.01,0.99)
        	{\includegraphics[scale = 0.25]{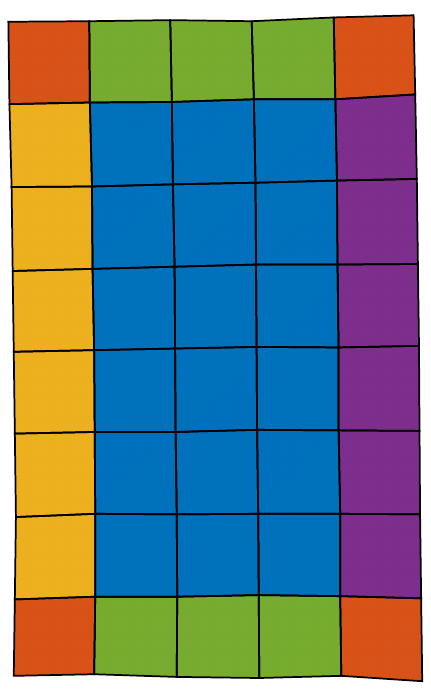}};  
    \end{axis}
\end{tikzpicture}}{(e) $T_2$}&
\subf{\includegraphics[height=\figHeight]{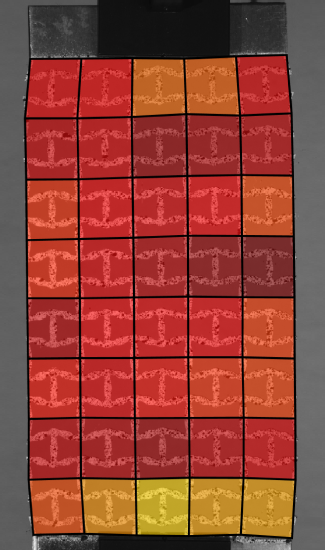}\\
\includegraphics{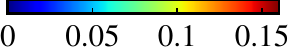}}{(f) $\varepsilon_{11}$}&
\subf{\includegraphics[height=\figHeight]{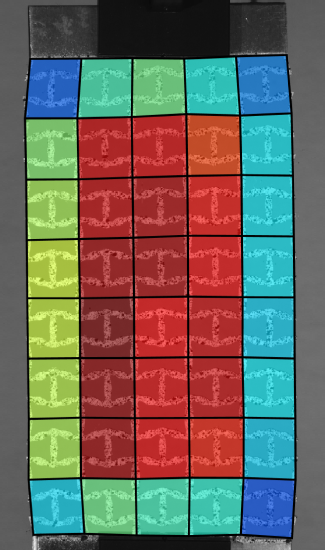}\\
\includegraphics{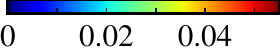}}{(g) $\varepsilon_{22}$}&
\subf{\includegraphics[height=\figHeight]{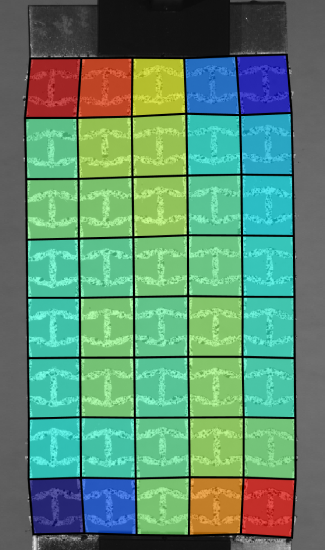}\\
\includegraphics{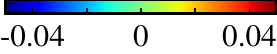}}{(h) $\varepsilon_{12}$}\\\\
\subf{
\begin{tikzpicture}[baseline]
    \begin{axis}[%
   		scale only axis,
		width=\figWidth, height=\figHeight, at={(0,0)},
		xlabel={Engineering shear $\gamma$},
		ylabel={Unit cell shear $\gamma$},
		xlabel near ticks,
		ylabel near ticks,
        grid=both,
		grid style={dotted, line width=.1pt, draw=black!50},
		enlargelimits=false,
		axis on top,
		clip=false
            ]
        \addplot[]
        	graphics[xmin=-0.45,xmax=0.05,ymin=-0.45,ymax=0.05]{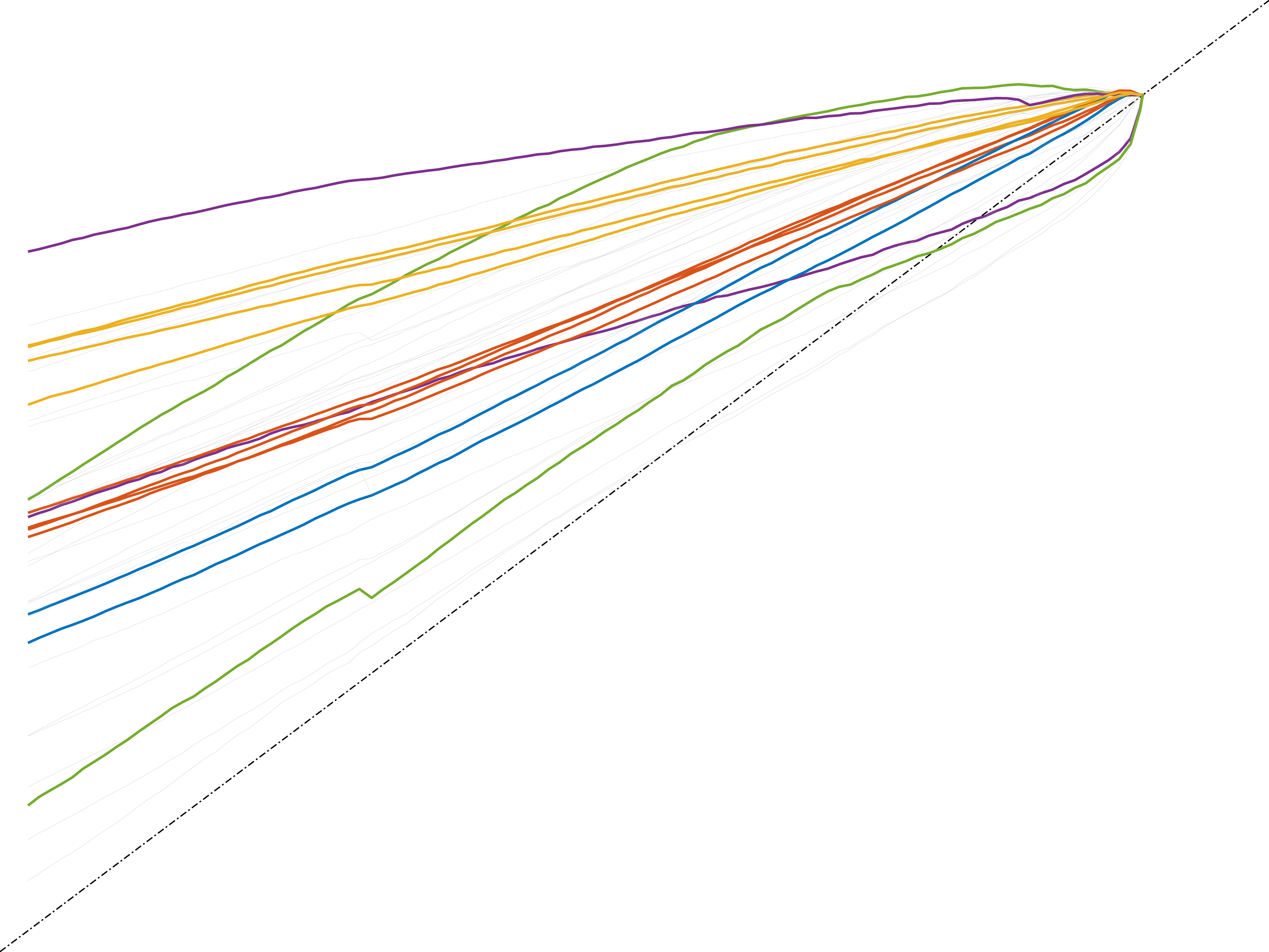};
        \node[above left] at (rel axis cs: 0.98,0.02) 
        	{\includegraphics[scale = 0.3]{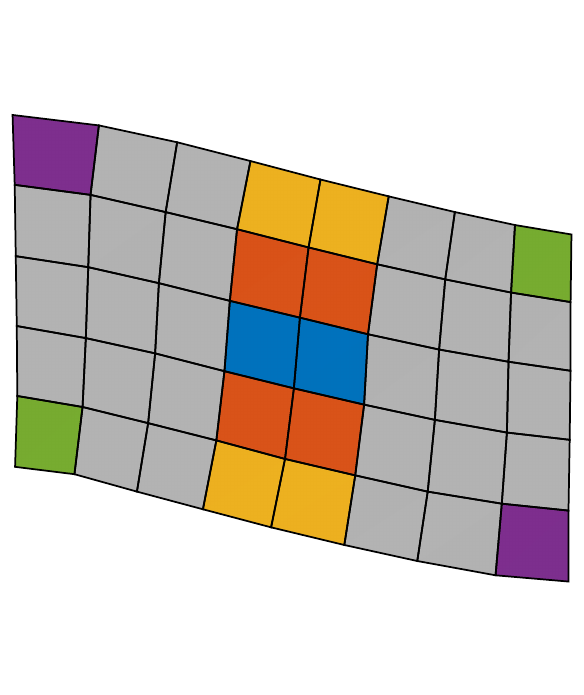}};
    \end{axis}   
\end{tikzpicture}}{(i) $S$}&
\subf{\includegraphics[height=\SfigHeight]{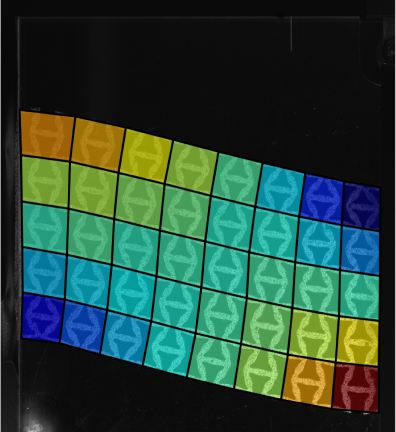}\\
\includegraphics{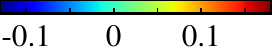}}{(j) $\varepsilon_{11}$}&
\subf{\includegraphics[height=\SfigHeight]{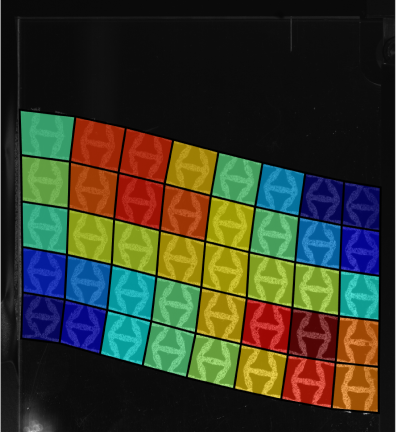}\\
\includegraphics{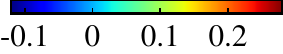}}{(k) $\varepsilon_{22}$}&
\subf{\includegraphics[height=\SfigHeight]{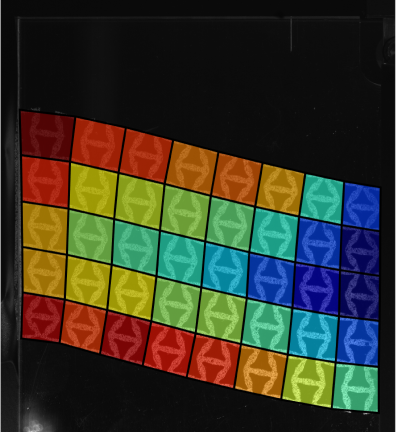}\\
\includegraphics{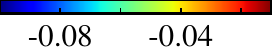}}{($\ell$) $\varepsilon_{12}$}
\end{tabularx}
\caption{Evolution of the macroscopic transverse strain $\varepsilon_{22}$ with respect to the longitudinal strain $\varepsilon_{11}$ for specimen $T_1$ (a) and specimen $T_2$ (e). Evolution of the macroscopic shear strain $\varepsilon_{12}$ with respect to the effective engineering shear strain $\gamma$ for specimen $S$ (i). The behaviour of the unit cells can be regrouped in bundles represented by different colours. Macroscopic strains maps obtained via \textit{local} DIC at a loading stage of 0.15 engineering strain, (b-d) for specimen $T_1$ ; (f-h) for specimen $T_2$ ; (j-$\ell$) for specimen $S$. }
\label{fig:strain_heterogeneity}
\end{figure*}

By looking at the deformed of the specimen $S$, we remark, more than in any other tests of the present work, a strong heterogeneity in the strain field. Rather than experiencing a homogeneous shear, the specimen $S$ undergoes rotations, leaving zones with predominant tension (top left and bottom right of $S$, see \autoref{fig:uniaxial_honeycomb}(k-$\ell$)), predominant compression (top right and bottom left of $S$), and predominant shear (at the center of $S$). These observations will be developed in the next paragraphs.

\paragraph{Scale of the unit cell} Next, we intend to analyse the global kinematics of the material, \textit{i.e.} the averaged kinematic values over the unit cells. To this end, we perform a local-DIC measurement for all the tests. We measure the macroscopic displacement at each node of the lattice, and derive the strain field, depicted in \autoref{fig:strain_heterogeneity}. In particular, \autoref{fig:strain_heterogeneity}(a,e) illustrate the evolution of the averaged transverse strain with respect to the averaged longitudinal strain for all unit cells of the specimens.  The ratio of the averaged strain components (\textit{i.e.} the slope of the curves) yields the effective Poisson's ratios, $\nu_{12}$  and $\nu_{21}$ respectively.\smallskip

At finite strains, the mechanical behaviour shifts rapidly, indicating in particular a decrease of the "auxeticity" of the specimen. Beyond 10\% effective strain, both effective Poisson's ratios no longer satisfy the small strain prediction of \cite{Agnelli2020} (reported also in \ref{appendix}). This effect is known in re-entrant honeycombs: the evolution of the Poisson's ratio with applied strain has already been observed and discussed in \cite{Wan2004}. Note also that improvements in the design of re-entrant honeycombs using a non-linear material behaviour in the optimization process would permit to stabilize the Poisson's ratio in a range up to 0.2 engineering strain, as shown in \cite{Zhang2019}.

\paragraph{Strain heterogeneity in the specimen} \label{sec:strain_heterogeneity}
We further explore the strain heterogeneity in the specimen. The question has an importance in itself, as mathematical optimisation methods are generally  defined on unit cells with periodic boundary conditions. Indeed, the interest is often on the \textit{macroscopic} behaviour of the structure, hence considered as a continuum material with \textit{homogenized} properties. The computation of this macroscopic apparent behaviour from the microscopic unit cell configuration (geometry and material properties) uses the assumption of an homogeneous state of strain in the structure \cite{Sanchez-Palencia1987}, equivalent to considering a specimen of infinite size. However, the specimen size is in practice limited by the experimental setup. As a consequence, boundary conditions applied to the specimen (free surfaces, clamping, etc.) are the source of strain heterogeneities.\smallskip

In all the tests, the macroscopic behaviour of the cells can be regrouped in \textit{bundles}, identified by curves with different colours in \autoref{fig:strain_heterogeneity}(a, e, i) . The scatter of the bundles is a evidence of heterogeneity in the specimen. For specimen $T_2$ (see \autoref{fig:strain_heterogeneity}(e)), there is merely a single line of cells which is affected by the boundary conditions, generally showing a lower transverse strain than center cells: cells associated to the clamped boundaries (in green and yellow) are constrained kinematically, while cells located on free edges (orange ad purple) are less strained transversely because of the vanishing transverse stresses. Apart from this boundary layer, the cells in the center of the specimen belong to the same bundle (coloured in blue), thus denoting a uniform state of strain in this region. Hence, the observed cell behaviour can be expected to be close to the homogenised behaviour; this is verified with the macroscopic Poisson's ratio identified close to the theoretical value of $\nu = -0.4$ (dash-dot black line).\smallskip

By opposition, the specimen $T_1$ (see \autoref{fig:strain_heterogeneity}(a)) shows an highly heterogeneous state of strain, with cell bundles that are more difficult to separate. This is mainly due to the higher absolute value of the Poisson's ratio ($\nu \approx -1.5$, dash-dot black line). At small strain \textit{i.e.} between 0 and 0.05 effective engineering strain, the specimen is rather homogeneous (besides the purple bundles, the unit cells all follow the same trend). Between 0.05 and 0.15, each bundle sequentially start to behave independently (yellow bundle, then green bundle, orange bundle, etc.). To better appreciate the average strain distribution in the specimen, A video of the test with the superimposed averaged strain field is provided (see \href{run:Movie3_T1_UnitCells_GL_Strains.mp4}{Movie 3}). We remark that at 0.15 effective engineering strain, we need three lines of cells from the constrained zones to neglect the influence of the boundary conditions. Hence, only the two central lines of the specimen are not affected by the boundary conditions (see \autoref{fig:strain_heterogeneity}(c)).\smallskip

Regarding the shear specimen S (see \autoref{fig:uniaxial_honeycomb}(i- $\ell$)), we notice that the unit cells shear strain $\gamma$ is in general lower than the engineering shear $\gamma_S$ imposed on the specimen. This is mostly due to the rotation of cells in the center region. In addition, a shear strain gradient is observed in the specimen, with a higher value in the center cells (in blue) that decreases with approaching boundaries (orange and yellow); this is in agreement with the free edge condition at which the shear stresses vanish. Moreover, the corner cells can be separated in two cases. First, bottom-left and top-right cells, in green, are first compressed in the early stages up to a point where contact occurs between members ($\gamma_S \approx 15\%$); then these cells are submitted to more shear in the latter stages. Second, top-left and bottom-right cells, in purple, are mostly stretched because of the specimen curvature. Despite the observed strain heterogeneity, it can be seen that the two center cells in blue are loaded proportionally to the imposed shear (with $\gamma \approx 0.65 \gamma_S$).

\subsection{Truss-hinge equivalent kinematic model}
\begin{figure}
\centering
\subf{\includegraphics[height=0.35\columnwidth]{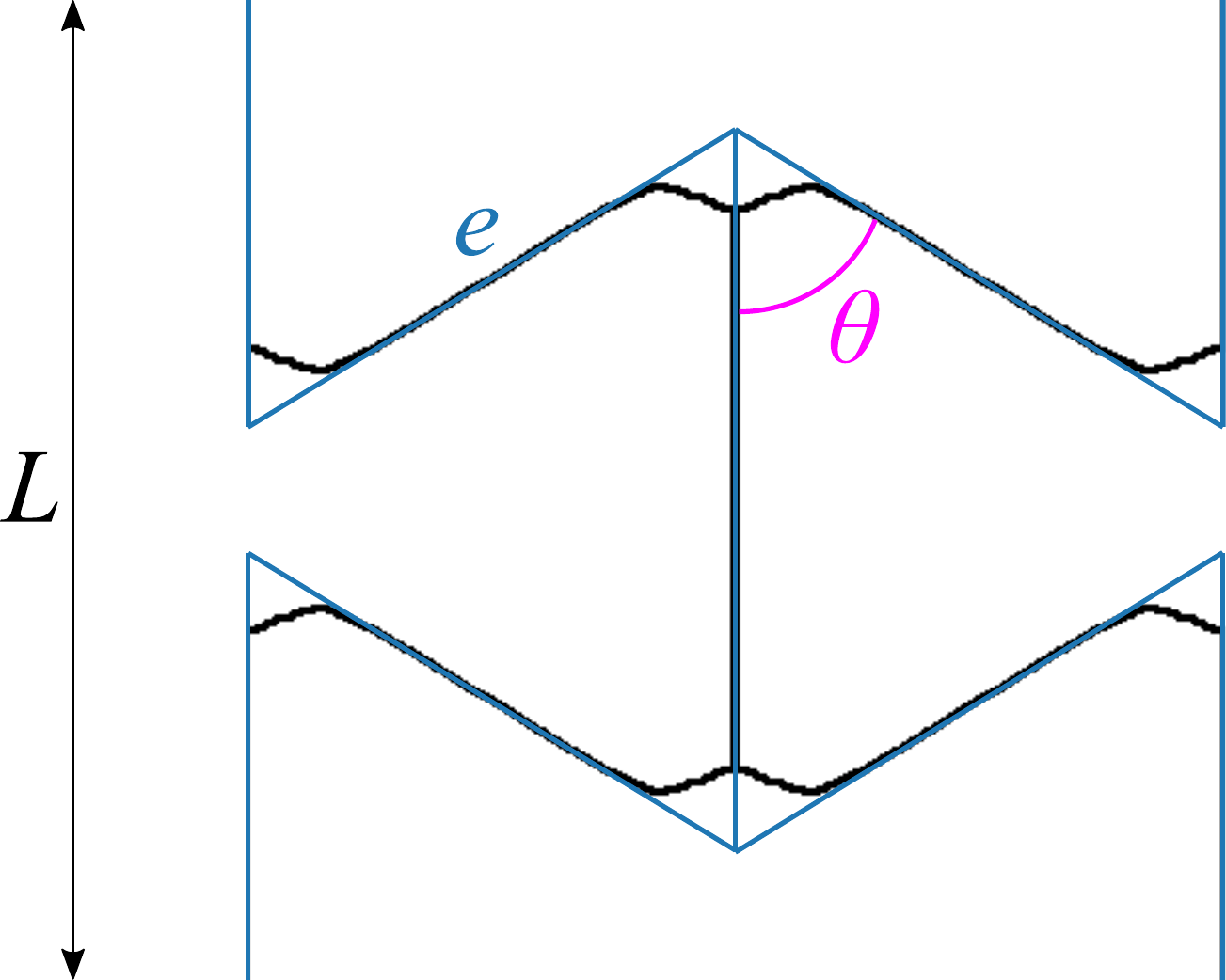}}{(a) $\mathcal{K}_{beams}\, (\theta_0 = 59^{\circ})$}
\hfill
\subf{\includegraphics[height=0.35\columnwidth]{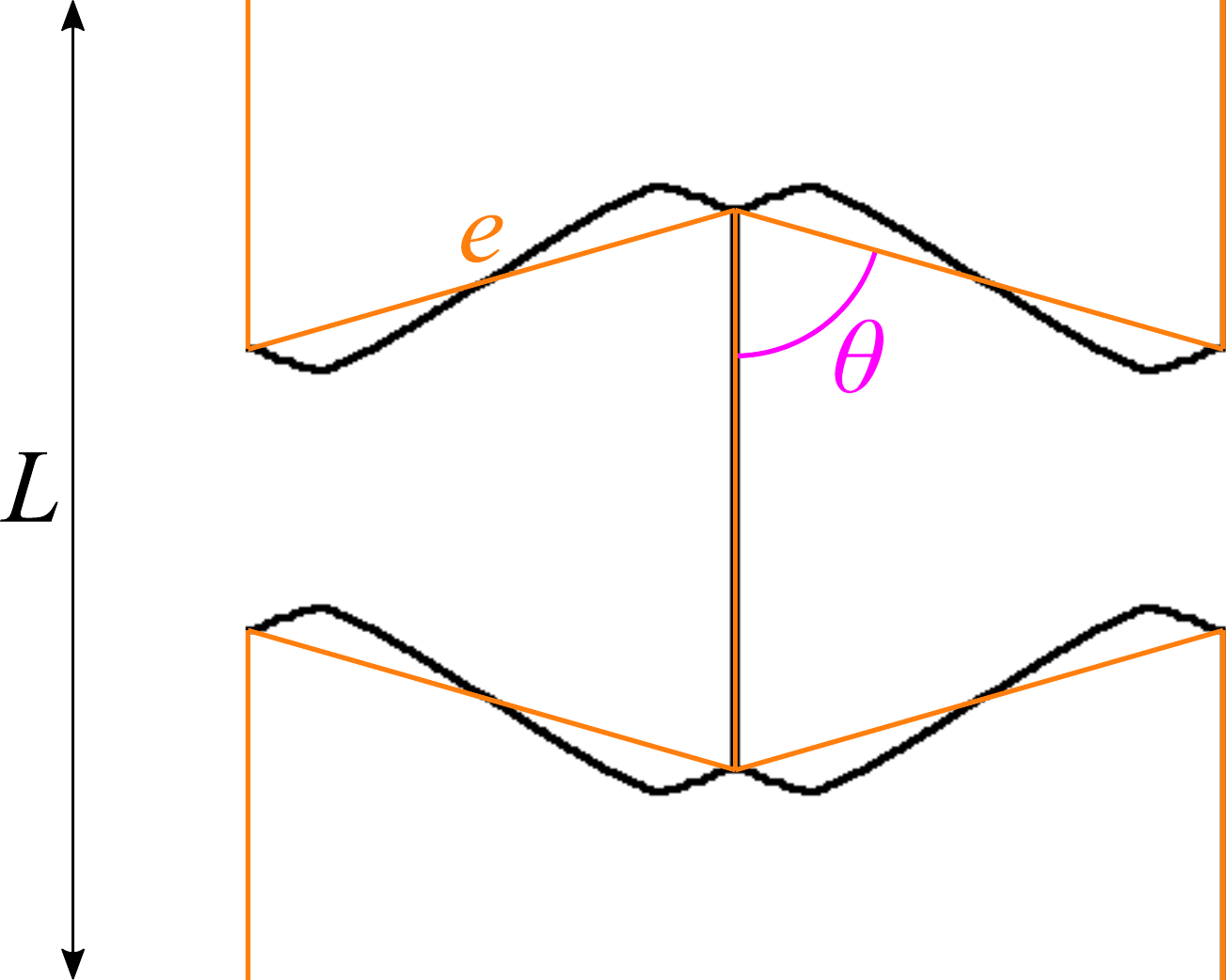}}{(b) $\mathcal{K}_{nodes}\, (\theta_0 = 74^{\circ})$}
\caption{Skeleton of the unit cell displayed in black with parametrization (a) $\mathcal{K}_{beams}$ and (b) using $\mathcal{K}_{nodes}$. }
\label{fig:skeleton}
\end{figure}

\begin{figure}[ht]
\centering
\subf{\includegraphics[width=0.85\columnwidth]{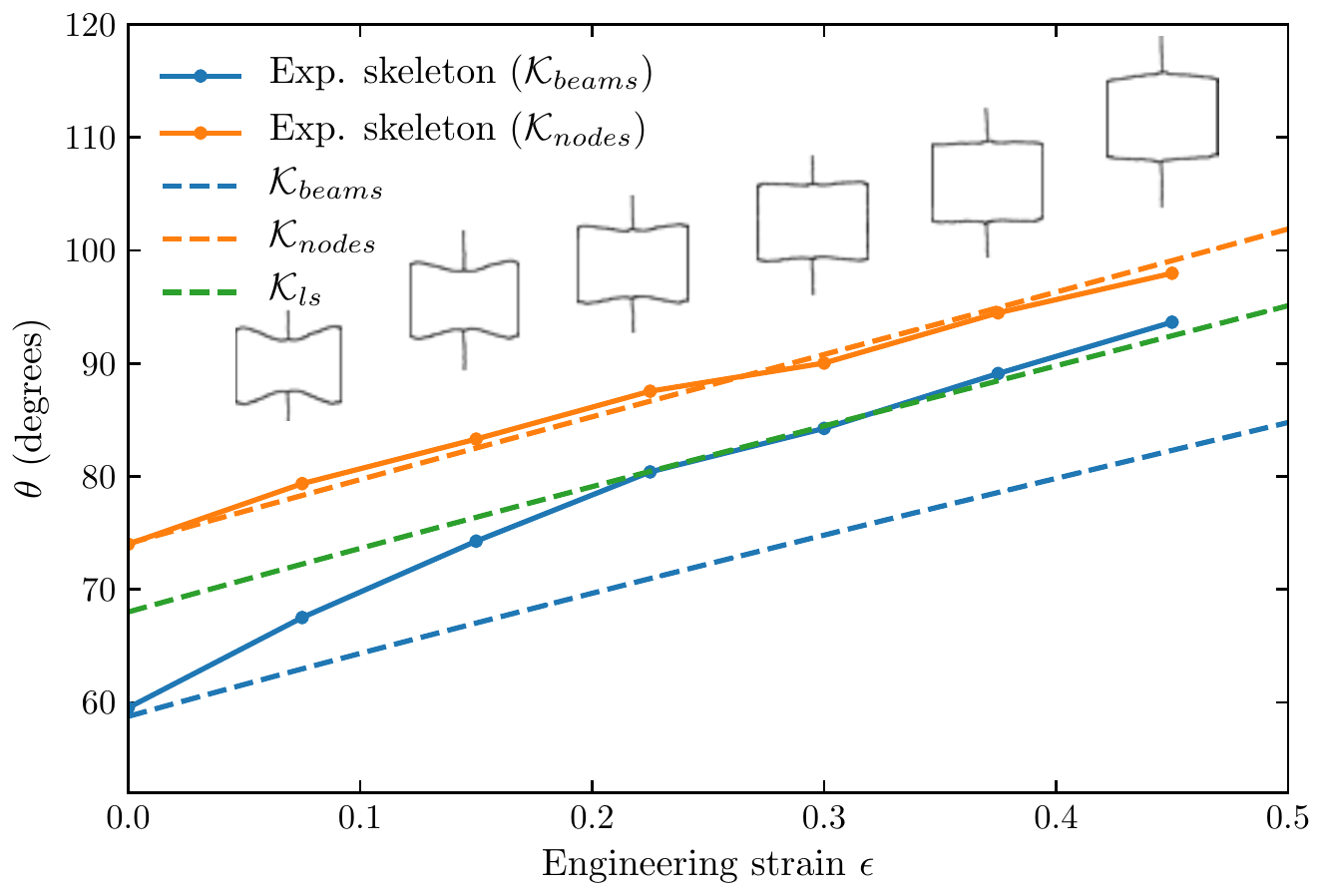}}{(a)}
\subf{\includegraphics[width=0.85\columnwidth]{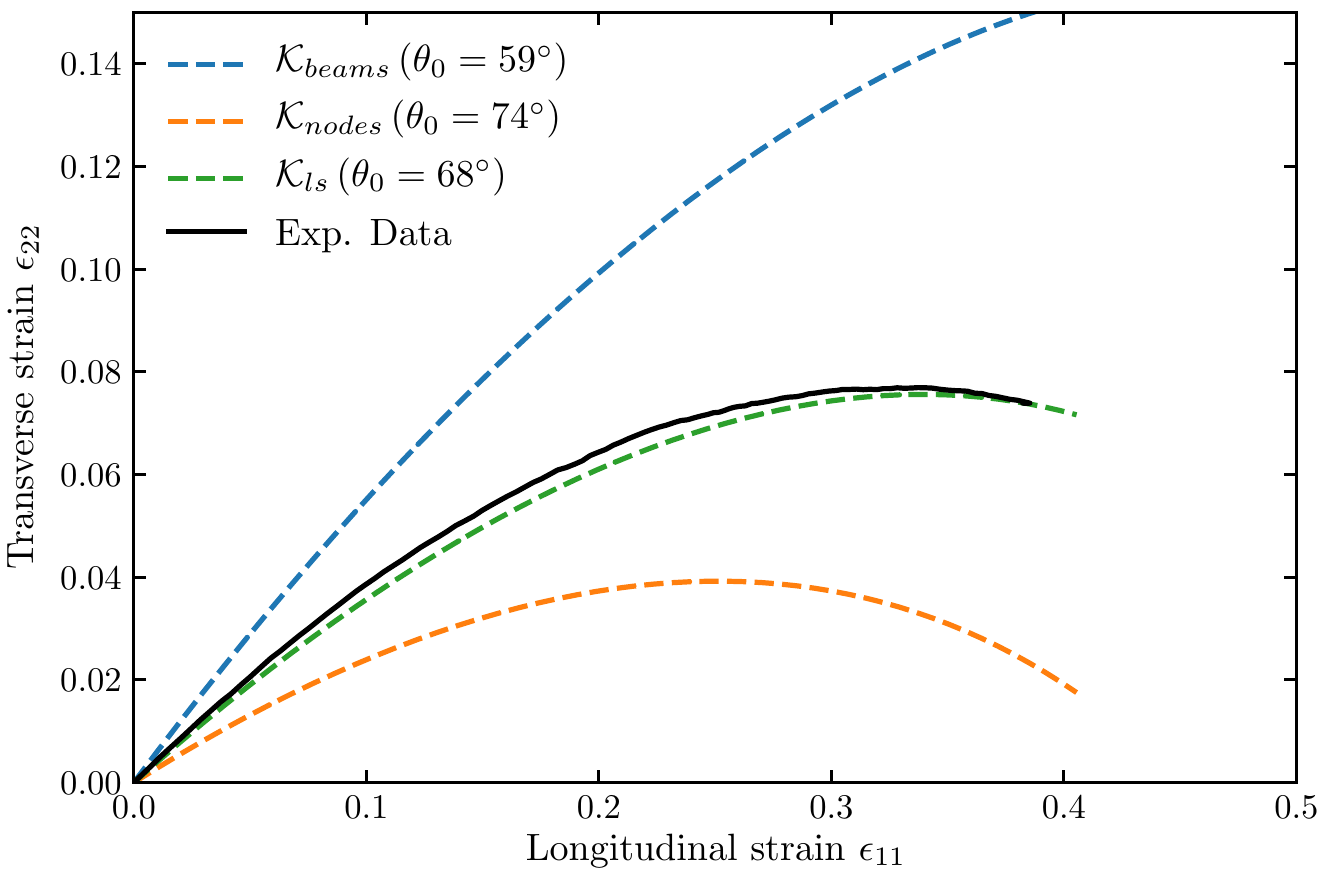}}{(b)}
\caption{(a) Evolution of $\theta$ with the engineering longitudinal strain. The experimental skeleton is computed on the central unit cell of the specimen from the pictures of the tensile tests at different strains. (b) Evolution of the transverse strain $\epsilon_{22}$ with respect to the longitudinal strain $\epsilon_{11}$. The experimental curve is obtained by computing an average of the curves belonging to the blue bundles in \autoref{fig:strain_heterogeneity}(e).}
\label{fig:trusses-hinges-kinematics}
\end{figure}

Since the strain distribution of specimen $T_2$ is localized at the hinges of the structure, we intend to examine whether a simple kinematic model with rigid trusses and rotating hinges (nodes) is sufficient to predict the Poisson's ratio of the structure. To this end, we derive a generic parametrization of the unit cell of \autoref{fig:new_micro} based upon its morphological \textit{skeleton}, which is a ``wire'' version of the shape that is equidistant to its boundaries. In shape ahnalysis, the \textit{skeleton} is frequently used as shape descriptors as it usually emphasizes geometrical and topological properties of the shape, such as its connectivity, topology, length, direction, and width. Interested readers may refer to \cite{Montanari1968, Kimmel1995} for a matematical definition of skeletons and algorithms to compute them. In our work, the morphological skeleton of our architecture is computed from a rasterized binary version of \autoref{fig:new_micro} via the \texttt{SkeletonTransform} command from Wolfram Mathematica (version 11.2, 2018). The obtained result is depicted in \autoref{fig:skeleton} (geometry in black). We remark that in spite of the relative complexity of the cell geometry, the corresponding skeleton can be decomposed in a reduced number of straight features (beams) and nodes connecting them (hinges). In order to model our structure as a simple re-entrant honeycomb, two configurations may be chosen:
\begin{itemize}[leftmargin =*]
\item configuration $\mathcal{K}_{beams}$ (depicted in blue in \autoref{fig:trusses-hinges-kinematics}(a)) is meant to emphasize the arrangement of the principal beams. The identification of the beams is easily achieved through a linear fit. The \texttt{ImageLines} command from Wolfram Mathematica finds line segments of a rasterized binary image and returns the coordinates of their endpoints. This configuration presumably yields the \textit{smallest} angle $\theta$.
\item configuration $\mathcal{K}_{nodes}$ (depicted in orange in \autoref{fig:trusses-hinges-kinematics}(b)) is meant to emphasize the position of the nodes. The identification of the nodes is done manually on the skeleton. This configuration presumably yields the \textit{largest} angle $\theta$.
\end{itemize}
Naturally, the real configuration may stand between $\mathcal{K}_{beams}$ and $\mathcal{K}_{nodes}$. This configuration should accurately predict the evolution of the effective transverse strain $\varepsilon_{22}$ with respect to the effective longitudinal strain $\varepsilon_{11}$ observed experimentally:
\begin{itemize}[leftmargin =*]
\item configuration $\mathcal{K}_{ls}$ is obtained by finding the angle $\theta$ which best fits the experimental experimental curve $\varepsilon_{22} = f(\varepsilon_{11})$. We use the least square method to find the best angle $\theta$ that fits the experimental curve. 
\end{itemize}
Given the equivalent truss-hinge model, we understand the whole unit cell kinematics are merely driven by the only variable angle $\theta$, therefore strain components can be expressed as:
\begin{equation}
\begin{aligned}
\text{Longitudinal:} \quad &\epsilon_{22}(\theta) = \frac{2 e}{L} \left({\cos(\theta_0) - \cos(\theta)}\right) \\
\text{Transverse:} \quad &\epsilon_{11}(\theta) = \frac{\sin(\theta)}{\sin(\theta_0)} - 1
\end{aligned}
\label{eq:truss_hinge_kinematics_E11}
\end{equation}
where $L$ is the characteristic length of the unit cell and $\theta_0$ denotes the initial value of $\theta$ (when the structure has not been stretched yet).\smallskip

Starting from the images of specimen $T_2$ recorded during the tensile test, we compute the morphological skeleton of the central unit cell and inferred a measure of the angle $\theta$ for both $\mathcal{K}_{beams}$ and $\mathcal{K}_{nodes}$. The evolution of $\theta$ measured during the experiments is compared to the rigid trusses rotating hinges model in \autoref{fig:trusses-hinges-kinematics}(a) for both $\mathcal{K}_{beams}$ and $\mathcal{K}_{nodes}$ skeletons. We remark that geometry $\mathcal{K}_{nodes}$ yield excellent agreement between model and experiments. Conversely, the model using configuration $\mathcal{K}_{beams}$ tends to underestimate the experiments. In addition, we plot the evolution of the transverse strain with respect to the longitudinal strain. We remark that the experimental evolution is bounded between the two configurations of the theoretical model $\mathcal{K}_{beams}$ and $\mathcal{K}_{nodes}$. Remarkably, we can identify an angle $\theta_0 = 68^\circ$ for which the theoretical kinematic evolution (equation \eqref{eq:truss_hinge_kinematics_E11}) is in good agreement with the experiments. It is worth noting that the $\theta_0 = 68^\circ$ case fits particularly well the end of the experimental blue curve of \Cref{fig:trusses-hinges-kinematics}(a). The obtained results support the idea that a rigid trusses rotating hinges kinematic model is suitable to predict the deformation pattern of specimen $T_2$ in spite of the soft elastomer used in the fabrication of the specimens.

\section{Concluding remarks} \label{sec:conclusion}
In this work, we have introduced a multi-scale experimental analysis designed to completely characterize the behaviour of architectured sheets undergoing extreme deformation. Our techniques have been applied to the analysis of a soft, auxetic sheet subjected to large tensile and shear loads (up to 0.5 effective strain). Based on this analysis, we are able to:
\begin{itemize}[leftmargin =*]
\item gain insight on the strain distribution of the specimen and identify the zones that have uniform strain field. This identification is particularly simple in our study, owing to our reconstruction of the macroscopic strain (the averaged kinematic values over each unit cell);
\item determine that strain heterogeneities dominate the response of finite-size specimens and that, to accurately capture the tensile response of an infinite sheet, the number of unit cells should be greater than four in both horizontal and transverse directions;
\item use the wealth of information obtained from the experiments to create a reduced order model (featuring rigid trusses and flexible hinges) that accurately describes the kinematic behaviour under tensile loads;
\item determine that, despite the strong heterogeneity displayed by the shear test results, it is possible to identify zones in the center of the specimen where the shear state is proportional to the applied engineering shear strain.
\end{itemize}
The tools presented in this study can be readily adapted to any two-dimensional architectured solid undergoing small or large deformations. In turn, the results that can be obtained by using these methods can potentially be leveraged to create tunable and stretchable mechanical devices \cite{Jiang2018a, Lee2019}.

\section*{Acknowledgements}
\noindent
This work was partly financed by the french-swiss ANR-SNF project MechNanoTruss (ANR-15-CE29-0024-01). F.A. acknowledges the support of the French doctoral fellowship ``Contrat Doctoral Spécifique pour Normalien''. C.D. acknowledges support from the US Army Research Office Grant W911NF- 17-1-0147.
\section*{Additional information}
\noindent
Supplementary material is available for this paper.
\begin{itemize}
\item Movie 1: Green-Lagrange strain field obtained via \textit{global} DIC at the center of specimen $T_1$;
\item Movie 2: Principal stretch $\lambda_1$ field obtained via \textit{global} DIC on specimen $T_2$;
\item Movie 3: Macroscopic strains maps obtained via \textit{local} DIC for specimen $T_1$.
\end{itemize}


\bibliographystyle{elsarticle-num}

\pagebreak
\appendix
\section{Small strain elasticity}
\label{appendix}
\paragraph{Orthotropic symmetry with 2-d linear elasticity} Let us denote by $Y$ the unit cell depicted in \autoref{fig:new_micro}(b). From a mechanical point of view, the equivalent homogeneous material displays an effective orthotropic behaviour. The linear elastic constitutive equation averaged over the unit cell relating the mean stress and strain tensors, denoted as $\vec{\sigma}^H$ and $\vec{\varepsilon}^H$ respectively, has the following expression for the two dimensional problems under consideration:
\begin{equation}
\begin{aligned}
\label{eq:sig:A:eps}
&\vec{\sigma}^H = \mathbb{C}^H \vec{\varepsilon}^H\\
\text{where:}\qquad &
\vec{\sigma}^H = \left \langle \vec{\sigma} \right \rangle_Y , \qquad
\vec{\varepsilon}^H = \left \langle \vec{\varepsilon} \right \rangle_Y.\\
& \mathbb{C}^H \, \text{is the homogenised stiffness tensor}
\end{aligned}
\end{equation}
In two-dimensional elasticity, the components of $\mathbb{C}^H$ in matrix notation and in Cartesian coordinates read:
\begin{equation} 
\label{eq:sig:A:eps:coord}
\begin{pmatrix} 
\sigma^H_{11} \\[3pt]
\sigma^H_{22} \\[3pt]
\sigma^H_{12}  
\end{pmatrix}
=
\begin{pmatrix} 
C^H_{1111} & C^H_{1122} & 0 		 \\[3pt]
C^H_{1122} & C^H_{2222} & 0          \\[3pt]
0 		   & 0 			& C^H_{1212} \\[3pt]
\end{pmatrix}
\begin{pmatrix} 
\varepsilon^H_{11} \\[3pt]
\varepsilon^H_{22} \\[3pt]
2 \varepsilon^H_{12}  
\end{pmatrix}
\end{equation}
Alternatively, one could express the effective strain as a function of the effective stress with the following effective material tensor:
\begin{equation}
\begin{pmatrix} 
\varepsilon^H_{11} \\[3pt]
\varepsilon^H_{22} \\[3pt]
2 \varepsilon^H_{12}  
\end{pmatrix}
=
\begin{pmatrix} 
1/E_1 & -\nu_{12}/E_2 & 0 \\[3pt]
-\nu_{21}/E_1 & 1/E_2 & 0 \\[3pt]
0 & 0 & 1/G \\[3pt]
\end{pmatrix}
\begin{pmatrix} 
\sigma^H_{11} \\[3pt]
\sigma^H_{22} \\[3pt]
\sigma^H_{12}  
\end{pmatrix}
\end{equation}
where $E_i$ denote the homogenized Young moduli, $\nu_{ij}$ denote the Poisson's ratios and $G$ denotes the homogenized shear modulus. Let us further remark, that by symmetry of the elastic compliance matrix, the following ratios have to be equal:
\begin{equation}
\frac{\nu_{12}}{E_2} = \frac{\nu_{21}}{E_1}
\end{equation}
The elastic moduli, $C^H_{ijkl}$, can equally be expressed in terms of the compliance moduli, \textit{i.e.} Young moduli and Poisson's ratios: $C^H_{1111}= (1-{\nu_{12}\nu_{21}})^{-1}E_1$, $C^H_{2222}= (1-{\nu_{12}\nu_{21}})^{-1}E_2$, $C^H_{1122}=\nu_{21}(1-{\nu_{12}\nu_{21}})^{-1}E_1$, $C^H_{2211}=\nu_{12}(1-{\nu_{12}\nu_{21}})^{-1}E_2$ with $C^H_{1122}=C^H_{2211}$ as can be easily obtained from the inversion of the corresponding matrices. A simple calculation immediately yields:
\begin{equation}
\nu_{12}=\frac{C^H_{1122}}{C^H_{2222}} \text{ and } \nu_{21}=\frac{C^H_{1122}}{C^H_{1111}}.
\label{eq:APR_stiffness_comp}
\end{equation}
Moreover, the homogenized Poisson's ratio $\nu_{ij}$ are equally denoted \textit{effective} Poisson's ratio to highlight their reference to the homogenized unit cell. For example $\nu_{12}$ characterizes the contraction of the structure in the direction of $Oy$ axis when the cell stretched in the direction of $Ox$ axis and in general $\nu_{12} \ne \nu_{21}$. Note that if the micro-architecture of the unit cell were to obey ``cubic'' symmetry, we would have $C^H_{1111} = C^H_{2222}$ and we would trivially obtain that $E_1=E_2=E^*$ and $\nu_{12} = \nu_{21}=\nu^*$.\smallskip

\paragraph{Experimental identification of the elastic coefficients} Hereafter we provide the complete experimental measurement of the effective elastic stiffness tensor. Let us recall that the effective constitutive law \eqref{eq:sig:A:eps} or alternatively \eqref{eq:sig:A:eps:coord} is a linear relation between the components of the effective stress and strain, from which the elastic moduli could be identified by a least square fitting. The main difficulty is that only the effective strain, $\vec{\varepsilon^H}$, can be directly measured from the experiment, see for instance \autoref{fig:strain_heterogeneity}. However, as suggested in \cite{Rethore2017}, the effective stress $\vec{\sigma^H}$ can be numerically computed from the experimental applied forces if the geometry and the constitutive behaviour of the base material are validated. As a consequence, $\mathbb{C}^H$, the effective elastic tensor of the design phase is obtained as a linear fit from $\vec{\varepsilon^H}$ and $\vec{\sigma^H}$. The computation could be performed on several unit cells of the specimen, yet here we will merely report the behaviour of the central unit cell. In order to compare the values of the elasticity tensor $\mathbb{C}^H$ computed in the design phase we have non-dimensionalized the resultant forces.\smallskip

For the computations, the elastic moduli of the base material were fixed according to \cite{Agnelli2020} for comparison purposes. Hence, the base material was defined with a Young's modulus $E_m = 0.91 \mathrm{MPa}$ and with a Poisson's ratio $\nu = 0.3$. Under the plane stress assumption, the components of the elastic tensor of the base material become $C^m_{1111} = C_m^{2222} = 1.0 \mathrm{MPa}$; $C_m^{1122} = 0.3 \mathrm{MPa}$ ; $C_m^{1212} = 0.35 \mathrm{MPa}$.

Experimentally, we remark that that $T_1$ is around four times stiffer than $T_2$ for a effective ranging from 0\% to 10\%.
\begin{table}[ht]
\centering
\begin{tabularx}{\columnwidth}{*{2}{>{\centering \arraybackslash}X}}
$\mathbb{C}^H(\omega)$ & $\mathbb{C}^{H,exp}(\omega)$\\
\hline
\vspace{0.3cm}
{$\displaystyle
	\begin{pmatrix}
    	0.12 & -0.05 & 0 \\
    	-0.05 & 0.04 & 0 \\
   		0 & 0 & 0.006 \\
  	\end{pmatrix}$}
&
\vspace{0.3cm}
{$\displaystyle
	\begin{pmatrix}
    	 0.1207 & -0.0487 & 0      \\
   		-0.0487 &  0.0318 & 0      \\
   		0       &  0      & 0.0044 \\
  	\end{pmatrix}$}
\end{tabularx}
\caption{Comparison between the effective $\mathbb{C}^H(\omega)$ (see also Table 1 of \protect \cite{Agnelli2020}) and measured elasticity tensor $\mathbb{C}^{H,exp}(\omega)$  displayed in the left and right column respectively. The measured elasticity tensor $\mathbb{C}^{H,exp}(\omega)$ was determined by combining DIC measurements and FEM computations.}
\label{table:meaneps}
\end{table}
\end{document}